\documentclass[twocolumn]{aastex631}

\usepackage{amsmath} 

\usepackage{color}

\usepackage{savesym}
\savesymbol{tablenum}
\usepackage{siunitx}
\restoresymbol{SIX}{tablenum}
\usepackage{booktabs}

\begin{document}

\title{The Atacama Cosmology Telescope: High-redshift measurement of structure growth from the cross-correlation of Quaia quasars and CMB lensing from ACT DR6 and \textit{Planck} PR4}

\author[0009-0001-3987-7104]{Carmen~Embil Villagra}\affiliation{DAMTP, Centre for Mathematical Sciences, University of Cambridge, Wilberforce Road, Cambridge CB3 OWA, UK}
\affiliation{Kavli Institute for Cosmology Cambridge, Madingley Road, Cambridge CB3 0HA, UK}

\author{Gerrit~Farren}\affiliation{Lawrence Berkeley National Laboratory, 1 Cyclotron Road, Berkeley, CA 94720, USA}

\author[0000-0002-3255-4695]{Giulio~Fabbian}\affiliation{Universit\'e Paris-Saclay, CNRS, Institut d'Astrophysique Spatiale, 91405, Orsay, France}
\affiliation{School of Physics and Astronomy, Cardiff University, The Parade, Cardiff, Wales CF24 3AA, United Kingdom}
\affiliation{Kavli Institute for Cosmology Cambridge, Madingley Road, Cambridge CB3 0HA, UK}

\author[0000-0003-4922-7401]{Boris~Bolliet}\affiliation{Department of Physics, University of Cambridge, Cambridge, United Kingdom}
\affiliation{Kavli Institute for Cosmology Cambridge, Madingley Road, Cambridge CB3 0HA, UK}

\author[0000-0003-3230-4589]{Irene Abril-Cabezas}\affiliation{DAMTP, Centre for Mathematical Sciences, University of Cambridge, Wilberforce Road, Cambridge CB3 OWA, UK}
\affiliation{Kavli Institute for Cosmology Cambridge, Madingley Road, Cambridge CB3 0HA, UK}

\author[0000-0002-4598-9719]{David~Alonso}
\affiliation{Department of Physics, University of Oxford, Denys Wilkinson Building, Keble Road, Oxford OX1 3RH, United Kingdom}

\author[0000-0003-3479-7823]{Anthony~Challinor}
\affiliation{Institute of Astronomy, Madingley Road, Cambridge CB3 0HA, UK}
\affiliation{DAMTP, Centre for Mathematical Sciences, University of Cambridge, Wilberforce Road, Cambridge CB3 OWA, UK}
\affiliation{Kavli Institute for Cosmology Cambridge, Madingley Road, Cambridge CB3 0HA, UK}

\author[0000-0002-7450-2586]{Jo~Dunkley} 
\affiliation{Joseph Henry Laboratories of Physics, Jadwin Hall, Princeton University, Princeton, NJ, USA 08544} 
\affiliation{Department of Astrophysical Sciences, Peyton Hall, Princeton University, Princeton, NJ USA 08544}

\author[0000-0002-0935-3270]{Joshua~Kim}
\affiliation{Department of Physics and Astronomy, University of Pennsylvania, Philadelphia, PA 19104, USA}

\author{Niall~MacCrann} 
\affiliation{DAMTP, Centre for Mathematical Sciences, University of Cambridge, Wilberforce Road, Cambridge CB3 OWA, UK} \affiliation{Kavli Institute for Cosmology Cambridge, Madingley Road, Cambridge CB3 0HA, UK}

\author[0000-0002-5389-3565]{Fiona~McCarthy}
\affiliation{DAMTP, Centre for Mathematical Sciences, University of Cambridge, Wilberforce Road, Cambridge CB3 OWA, UK}
\affiliation{Kavli Institute for Cosmology Cambridge, Madingley Road, Cambridge CB3 0HA, UK}
\affiliation{Center for Computational Astrophysics, Flatiron Institute, New York, NY, USA 10010}

\author{Kavilan~Moodley}
\affiliation{Astrophysics Research Centre, University of KwaZulu-Natal, Westville Campus, Durban 4041, South Africa}
\affiliation{School of Mathematics, Statistics \& Computer Science, University of KwaZulu-Natal, Westville Campus, Durban 4041, South Africa}

\author[0000-0001-7805-1068]{Frank J.~Qu}
\affiliation{Kavli Institute for Particle Astrophysics and Cosmology, Stanford University, 452 Lomita Mall, Stanford, CA, 94305, USA}
\affiliation{Department of Physics, Stanford University, 382 Via Pueblo Mall, Stanford, CA, 94305, USA}

\author{Blake~Sherwin}\affiliation{DAMTP, Centre for Mathematical Sciences, University of Cambridge, Wilberforce Road, Cambridge CB3 OWA, UK}
\affiliation{Kavli Institute for Cosmology Cambridge, Madingley Road, Cambridge CB3 0HA, UK}

\author[0000-0002-8149-1352]{Crist\'obal Sif\'on}\affiliation{Instituto de F\'isica, Pontificia Universidad Cat\'olica de Valpara\'iso, Casilla 4059, Valpara\'iso, Chile}

\author[0000-0002-3495-158X]{Alexander van Engelen} \affiliation{School of Earth and Space Exploration, Arizona State University, 781 Terrace Mall, Tempe, AZ 85287, U.S.A.}

\author[0000-0002-7567-4451]{Edward J.~Wollack}\affiliation{NASA Goddard Space Flight Center, 8800 Greenbelt Road, Greenbelt, MD 20771, USA}

\correspondingauthor{Carmen~Émbil Villagrá}
\email{ce425@cam.ac.uk}

\begin{abstract}

We measure the amplitude of matter fluctuations over a wide range of redshifts by combining CMB lensing observations from ACT DR6 and \textit{Planck} PR4 with the overdensity of quasars from Quaia, a \textit{Gaia} and \textit{unWISE} quasar catalog. Our analysis includes the CMB lensing power spectrum from ACT DR6, the auto-correlation of two Quaia quasar samples centered at $z \simeq 1.0$ and $z \simeq 2.1$, and their cross-correlations with CMB lensing from both ACT DR6 and \textit{Planck} PR4. 
By performing a series of contamination and systematic null tests, we find no evidence for contamination in the lensing maps, contrary to what was suggested in previous Quaia cross-correlation analyses using \textit{Planck} PR4 CMB lensing data. From the joint analysis of the quasar auto- and cross-correlations with CMB lensing, and including BOSS BAO data to break the degeneracy between  $\Omega_m$ and $\sigma_8$, we obtain $\sigma_8 = 0.802^{+0.045}_{-0.057}$, consistent with $\Lambda$CDM predictions from \textit{Planck} primary CMB measurements. Combining the CMB lensing auto-spectrum with the cross-correlation measurement improves the constraint on $\sigma_8$ by 12\% relative to the lensing auto-spectrum alone, yielding $\sigma_8 = 0.804 \pm 0.013$. This dataset combination also enables a reconstruction of structure growth across redshifts. We infer a 12\% constraint on the amplitude of matter fluctuations at $z > 3$, with a measurement at the median redshift of the signal of $\sigma_8(\tilde{z}=5.1) = 0.146^{+0.021}_{-0.014}$, consistent with \textit{Planck} at the $1.4\sigma$ level. These results provide one of the highest redshift constraints on the growth of structure to date.

\end{abstract}

\tableofcontents

\section{Introduction} \label{sec:intro}

Measuring the growth of cosmic structure across time offers a powerful probe of both galaxy formation and fundamental physics. While the $\Lambda$ cold dark matter ($\Lambda$CDM) model has proven remarkably successful in explaining a wide range of cosmological observations, several fundamental questions remain unanswered, such the sum of neutrino masses, the nature of dark energy, or the properties of dark matter. A key strategy for addressing some of these open questions is to constrain the amplitude of matter fluctuations, typically parameterized by $\sigma_8$ or the derived quantity $S_8 \equiv \sigma_8 (\Omega_m/0.3)^{0.5}$, through observations of large-scale structure (LSS) tracers.

Analyses combining galaxy clustering and weak lensing measurements from surveys such as the Kilo-Degree Survey \citep[KiDS;][]{Kuijken_2015,Heymans_2021}, Dark Energy Survey \citep[DES;][]{Flaugher_2015,Abbott_2022}, and the Hyper SuprimeCam \citep[HSC;][]{Aihara_2017,more2023hypersuprimecamyear3,miyatake2023hypersuprimecamyear3,Sugiyama_2023} have reported values of $S_8$ that lie 2--3$\sigma$ below those inferred from primary cosmic microwave background (CMB) measurements by \textit{Planck} \citep{Planck2020} or the Atacama Cosmology Telescope (ACT; \citealt{louis2025atacamacosmologytelescopedr6}). A subsequent reanalysis combining galaxy and CMB lensing data \citep{Garc_a_Garc_a_2021} also found a $\sim$3$\sigma$ discrepancy, primarily driven by galaxy weak lensing information at low redshift. However, updated measurements from KiDS \citep{wright2025kidslegacycosmologicalconstraintscosmic,stölzner2025kidslegacyconsistencycosmicshear} and DES \citep{descollaboration2025darkenergysurveyyear} have been found to be in agreement with primary CMB measurements, significantly alleviating the ``$S_8$ tension''.

In this context, gravitational lensing of the CMB (for a review, see, e.g.,~\citealt{LEWIS_2006}) constitutes a well-understood tracer of the total matter distribution. As CMB photons traverse the large-scale structure from the surface of last scattering to the present day, their paths are deflected by gravitational potentials induced by matter overdensities. Therefore, the lensing of the CMB is an unbiased, integrated probe of matter along the line of sight. Since it probes the same matter field as galaxy surveys, we can cross-correlate the lensing convergence field with the galaxy overdensity at localized redshifts to ``slice'' the CMB lensing information, a technique known as ``lensing tomography''. 

Measurements of the CMB lensing auto-correlation \citep{qu2023atacama,Madhavacheril_2024,qu2025unifiedconsistentstructuregrowth,Carron_2022,SPT_Muse}, as well as recent cross-correlations \citep{Farren_2024} with \textit{unWISE} galaxies \citep{Wright_2010, Meisner_2019} have shown good agreement with $\Lambda$CDM predictions from primary CMB measurements. Meanwhile cross-correlations with Dark Energy Spectroscopic Instrument \citep[DESI;][]{DESI_Collaboration_2022} Luminous Red Galaxies \citep[LRG;][]{Kim_2024,Sailer_2024} and the DESI Legacy Survey \citep{DESI_Legacy,Qu_DESI_2025} have reported mildly lower amplitudes of structure growth (at roughly the $2\sigma$ level) relative to \textit{Planck}. However, more recent measurements at lower redshift, using the DESI Bright Galaxy Survey~\citep[BGS;][]{Hahn_2023} in cross-correlation with CMB lensing from both \textit{Planck} and ACT, show good consistency with primary CMB constraints \citep{Sailer:2025rks}.

Quasars offer a unique opportunity in this context \citep{Sherwin_2012,Geach_2013,Geach_2019,debelsunce2025cosmologyplanckcmblensing}, as they trace the large-scale structure at higher redshifts than conventional galaxy surveys. In particular, Quaia \citep{Storey_Fisher_2024} is the quasar catalog covering one of the largest comoving volumes compared to any previous spectroscopically calibrated quasar sample \citep{Lyke_2020}. This makes Quaia an ideal sample for cross-correlation with CMB lensing to probe the growth of structure at remarkably high redshifts ($z \lesssim 3$) and linear scales.

The cross-correlation of Quaia with CMB lensing was previously studied in \citet{alonso2023} using \textit{Planck} PR4 lensing maps, yielding a constraint of $\sigma_8 = 0.766 \pm 0.034$, consistent with primary CMB measurements from \textit{Planck} at the 1.4$\sigma$ level. This slightly low value was found to be driven by the highest redshift Quaia bin ($z \sim 2.1$) and repeating the analysis using polarization-only lensing maps gave a higher $\sigma_8$ value more consistent with \textit{Planck}, at the 0.2$\sigma$ level. This raised concerns about possible contamination in the \textit{Planck} PR4 lensing temperature reconstruction, particularly from high-redshift extragalactic foregrounds such as the cosmic infrared background (CIB).

A follow-up study \citep{Piccirilli:2024xgo} explored this further by splitting the quasar catalog into narrower redshift bins and localizing a mild discrepancy around $z \sim 1.7$; however, no conclusive evidence for CIB or any other extragalactic source of contamination was found to explain this discrepancy. In this work, we revisit the measurement using \textit{Planck} PR4 lensing maps \citep{Carron_2022}, adopting a more conservative set of analysis choices discussed in Appendix \ref{app:Planck-reanalysis}, and extend the study by incorporating measurements from the ACT sixth data release (DR6) CMB lensing maps \citep{qu2023atacama}. For the latter, we perform a series of systematics and contaminant tests, which will address whether this discrepancy arises due to contamination in the CMB lensing maps, residual systematics in either dataset, or a statistical fluctuation. 

Moreover, given the high-redshift sensitivity of the Quaia quasars \citep{Storey_Fisher_2024}, we extend the analysis beyond the traditional combination of the galaxy auto-power spectrum and the galaxy-CMB lensing cross-correlation by further including the CMB lensing power spectrum from ACT DR6 \citep{qu2023atacama,Madhavacheril_2024}. Adopting the terminology standard in galaxy weak lensing analyses, we refer to the combination of $C_\ell^{gg}$ and $C_\ell^{\kappa g}$ as a `2$\times$2pt' analysis, and its extension including the lensing auto-spectrum $C_\ell^{\kappa\kappa}$ as `3$\times$2pt'. This joint analysis probes the growth of matter fluctuations at redshifts beyond those directly covered by the quasars. By parameterizing the amplitude of the linear matter power spectrum in different redshift intervals, we reconstruct $\sigma_8(z)$ and obtain one of highest redshift measurements of structure growth ($z > 3$) to date.

The paper is organized as follows. In Section~\ref{sec:data}, we describe the datasets used in this analysis. Section~\ref{sec:theory} outlines the theoretical modeling of the cross-correlation measurement, as well as the redshift-evolution parametrization of the matter power spectrum used to reconstruct $\sigma_8(z)$. The measurement of the 2$\times$2pt spectra, along with the simulations, transfer functions, and covariance matrices, used in our analyses is presented in Section~\ref{sec:tomography-measurement}. In Section~\ref{sec:tests}, we detail a suite of systematic and contamination tests. The blinding policy, likelihood, and choice of priors are discussed in Section~\ref{sec:cosmoanal}. We then present and discuss the cosmological results from both the 2$\times$2pt and 3$\times$2pt analyses in Section~\ref{sec:results}. Finally, we summarize our main findings and conclusions in Section~\ref{sec:conclussion}. A series of appendices provide further details of our reanalysis of Quaia quasars with \textit{Planck} PR4 lensing, and the generation of mock quasar catalogs that we use in our tests for extragalactic foreground contamination.

\section{The Data}\label{sec:data}

\subsection{Quaia: the \textit{Gaia--unWISE} quasar catalog} \label{sec:data-Quaia}

Quaia is an all-sky quasar catalog obtained with the combination of quasar candidates from \textit{Gaia}'s third data release \citep{Gaia_2023} and infrared photometry from the \textit{unWISE} reprocessing of the Wide-Field Infrared Survey Explorer~\citep[WISE;][]{Wright_2010, Meisner_2019}. This combination allows for the initial homogeneous and complete \textit{Gaia} catalog to be used for cosmological and astrophysical purposes by improving the quality of the redshift estimates and purity of the sample when combining it with \textit{unWISE} data. 

The catalog includes just below 1.3 million sources of magnitudes $G<20.5$ and it is described together with a series of selection functions fitted to different redshift bins, with more detail in \cite{Storey_Fisher_2024}. These mitigate selection effects and are used as weighted masks in our cross-correlation analysis as discussed below. They account for dust and the scan patterns and source density of the parent surveys. We use an updated version of the catalog, with minor modifications described in \cite{Piccirilli:2024xgo}.\footnote{The catalog and selection functions are publicly available at \url{https://zenodo.org/records/8060755}.}

The redshift estimation of the sources uses a $k$-nearest neighbors ($k$NN) model with cross-matched Sloan Digital Sky Survey (SDSS) DR16Q quasars \citep{Lyke_2020} as labels and the \textit{Gaia} and \textit{unWISE} photometry, and \textit{Gaia}-estimated spectral redshift, as features. The final redshift is taken to be the median of the redshift of the nearest $k=27$ neighbors and the uncertainty $\sigma_z$ to be their standard deviation. 

We split the resulting catalog into two redshift bins at the median redshift of the full sample, defining Bin 1 as $z < 1.47$ and Bin 2 as $z > 1.47$. The objects in each bin correspond to those used to fit the selection functions described above. A summary of the key properties of each of the samples restricted to the ACT footprint can be found in Table \ref{table:Quaia}.

\begin{table}[]
    \centering
    \begin{tabular}{cccccc}
    \toprule
    Sample & $\bar{z}$ & $10^5$ SN & $\bar{n}[\text{deg}^{-2}]$ & $\ell_{\text{max}}$ & $b_g^i$ \\ \hline
    $z_1$  & 1.0       & 1.38      &   34.2    & 316   & $1.16^{+0.22}_{-0.25}$    \\
    $z_2$  & 2.1       & 1.29      &   37.7    & 526   & $1.02^{+0.20}_{-0.23}$   \\ 
    \bottomrule
    \end{tabular}
    \caption{Summary of the main properties of the two Quaia samples, restricted to the ACT footprint, which are used in this analysis. The mean redshift $\bar{z}$ is computed as the weighted average of the $dN/dz$ distribution for each bin. The quoted shot noise (SN) values are obtained by deconvolving the catalog measurement using Eq.~\eqref{eq:sn}. The maximum multipole $\ell_{\text{max}}$ indicates the largest included in the analysis for each bin; see Section~\ref{sec:measurement} for a discussion of the scale cuts. Finally, the $b_g^i$ parameters are presented in Equation \eqref{eq:bias}, and measured with the data combination: ACT$\times$Quaia + BAO, for which cosmological results are presented in Section \ref{sec:results-2x2pt-ACT}.}
    \label{table:Quaia}
\end{table}

\begin{figure}
 \centering
 \includegraphics[width=\linewidth]{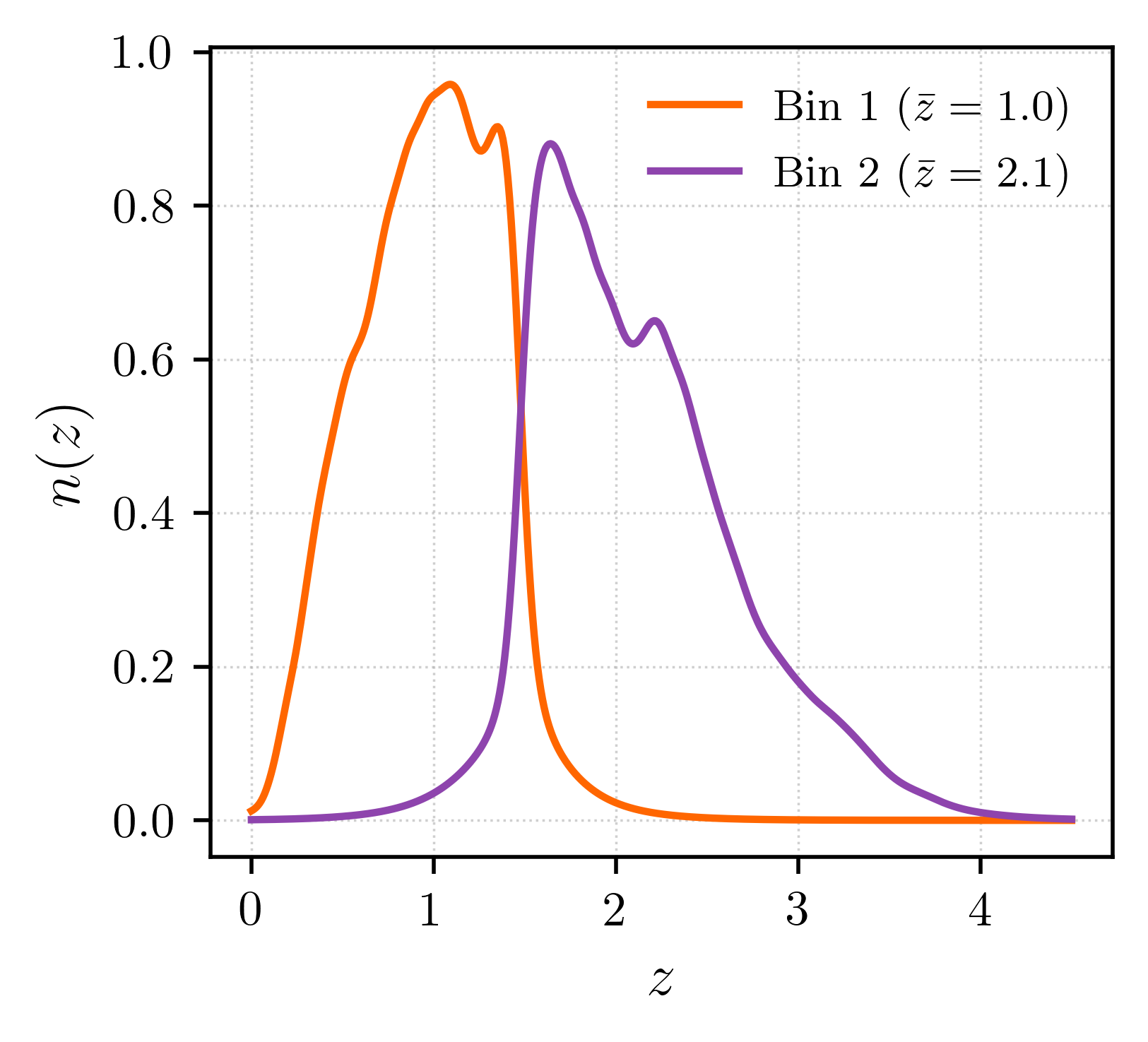}
 \caption{Normalized redshift distributions of the two Quaia samples used in our measurement with mean redshifts of $z\sim 1.0$ and $z\sim 2.1$ for Bin 1 and Bin 2, respectively. These distributions have been estimated using PDF stacking on quasars in the joint ACT--Quaia footprint of each redshift bin as described in the main text.} 
 \label{fig:dndz}
\end{figure}

The redshift distributions ($dN/dz$) of the two samples are needed to compute the theory predictions of our model as described in Section \ref{sec:theory}. A thorough investigation of Quaia's $dN/dz$ estimation was done in \cite{alonso2023}, finding that cosmological parameters were robust against mild mis-calibrations of the redshift distributions. In particular, it was shown that cosmological constraints remain ``virtually unchanged'' whether the redshift distribution is estimated using the direct calibration method of \citet{Lima_2008} or through probability-density-function (PDF) stacking. In this work, we adopt the latter approach to recompute the redshift distributions of the quasar samples on the ACT footprint (see Figure \ref{fig:dndz}). This method assumes a Gaussian error model, estimating the redshift distribution by stacking a Gaussian PDF $\mathcal{N}(z, \sigma_z)$ for each quasar, where $z$ is the redshift of the quasar and $\sigma_z$ its associated uncertainty. We find up to a 5\% difference compared to the redshift distributions computed over the full Quaia footprint, and therefore use the recomputed $dN/dz$ whenever predicting the theory for the ACT footprint.

We construct the quasar overdensity ($\delta_{\mathrm{g}}$ hereafter) maps in each redshift bin using \texttt{HEALPix} \citep{healpix} with a resolution parameter of $\texttt{nside} = 512$. The overdensity in each pixel $p$ is computed as
\begin{equation}
    \delta_{\mathrm{g},p} = \frac{N_p}{\bar{N}\omega_p} - 1 \; ,
\end{equation}
where $\omega_p$ is the value of the selection function in pixel $p$, $N_p$ is the number of quasars in each pixel, and $\bar{N}$ is the mean number density of quasars in the sample. We compute $\bar{N}$ accounting for the selection function, by
\begin{equation}
    \bar{N} = \frac{\sum_p N_p}{\sum_p \omega_p} \; .
    \label{eq:nbar}
\end{equation}
Moreover, $\omega_p$ is used as a weighted mask with a threshold of 0.5, masking all pixels with a value below this threshold to avoid instabilities where $\omega_p$ is close to zero.\footnote{It was shown in \cite{alonso2023} that cosmological parameters are robust against changes in this threshold.} Therefore, only unmasked pixels are used to calculate the mean density in \eqref{eq:nbar}. To compute the auto-correlation measurement on the ACT footprint we intersect the Quaia selection function with the ACT baseline mask described in Section \ref{sec:data-ACT} and perform a C1-type apodization of 0.2 degrees\footnote{We note that the Quaia footprints for both redshift bins contain nearby holes, so a broader apodization could bias the measurement by, e.g., apodizing a region common to two different holes twice. We have verified that our spectra are stable under reasonable changes in the apodization width.} using \texttt{NaMaster}\footnote{\url{https://github.com/LSSTDESC/NaMaster}} \citep{Alonso_2019} before computing the power spectra.

To model the shot-noise contribution to the quasar auto-spectra, we assume Poisson statistics and compute the shot noise analytically as
\begin{equation}
    \hat{N}_\ell = \frac{\left< \omega \right>_p}{\bar{N}_\Omega} \; ,
    \label{eq:sn}
\end{equation}
where $\left< \omega \right>_p$ denotes the mean value of the selection function in the analysis footprint and $\bar{N}_\Omega$ is the angular number density of quasars given by $\bar{N}_\Omega = \bar{N}/\Omega_{\text{pix}}$. Here, $\bar{N}$ is the mean number density given by Equation~\eqref{eq:nbar}, and $\Omega_{\rm pix}$ is the solid angle of each \texttt{HEALPix} pixel at $\texttt{nside} = 512$. In our baseline parameter analysis, we will adopt a Gaussian prior centered at the value given by $\hat{N}$ and with a 10\% width to account for a potential mismodeling of the shot noise.

\subsection{ACT DR6 CMB lensing} \label{sec:data-ACT}

The CMB lensing convergence map used as our baseline is reconstructed from the CMB temperature and polarization anisotropy data \citep{2003_Okamoto_hu,Maniyar_2021} from ACT DR6 \citep{Madhavacheril_2024, qu2023atacama,maccrann2023atacama}. It is produced using night-time-only CMB data collected between 2017 and 2021 in the $90\,\text{GHz}$ and $150\,\text{GHz}$ bands.

The DR6 lensing maps cover 9400 deg$^2$ of the sky and are signal dominated for multipoles $L<150$. They are reconstructed using a split-based quadratic estimator \citep{Madhavacheril_2021}, which uses split maps made from disjoint observations, as opposed to a single CMB map, to ensure that the instrumental and atmospheric noise of each map is independent. This makes the lensing bias subtractions -- the mean-field removed from the reconstructed lensing map and the Gaussian noise bias removed from the reconstructed power spectrum -- insensitive to the modeling of these noise components.

The ACT DR6 lensing maps are reconstructed using a quadratic estimator on CMB multipoles $600<\ell<3000$. The large scale cut $\ell<600$ is chosen to exclude the majority of atmospheric noise, Galactic foregrounds \citep{abrilcabezas2025impactgalacticnongaussianforegrounds}, and an instrument-related transfer function \citep{Naess_2023}; while small scales of $\ell>3000$ are excluded to minimise the impact of astrophysical contamination from the CIB, radio sources, and the thermal Sunyaev--Zeldovich (tSZ) effect \citep{1969Ap&SS...4..301Z,1970SZ} in the temperature maps \citep{maccrann2023atacama}. To mitigate further the effects of extragalactic foregrounds, the lensing maps are constructed using a profile-hardened lensing quadratic estimator \citep{Namikawa_2013, Osborne_2014, Sailer2020, Sailer2023}. This quadratic estimator is immune, at leading order, to the mode couplings coming from objects with radial profiles such as those expected from tSZ clusters. In Section \ref{sec:tests}, we  explore further the impact of extragalactic foregrounds, showing that they have a negligible impact in our cross-correlation measurement.

In addition to the cross-correlation of ACT DR6 lensing with the quasar overdensity, we also include the CMB lensing power spectrum to perform a 3$\times$2pt analysis. The computation of this auto-correlation measurement and its associated covariance matrix are described in detailed in \citet{qu2023atacama}.\footnote{Both these products are publicly available at \url{https://github.com/ACTCollaboration/act_dr6_lenslike}.} We use the baseline multipole range $40<L<763$ where the auto-spectrum is measured to 2.3\% precision, corresponding to a $43\sigma$ significance.

The baseline ACT DR6 lensing mask is constructed by intersecting the ACT footprint with the \textit{Planck} mask that selects the 60\% of the sky with the lowest dust contamination. This mask is then apodized using a cosine roll-off at the edges. A detailed description of this baseline mask can be found in \cite{qu2023atacama}. In Section \ref{sec:tests}, we perform a series of contaminant tests using different intersections with \textit{Planck} masks: one covering 40\% of the sky with lowest dust contamination and a different 60\% mask that removes dust clouds on the edges of the baseline mask, which is used for the CIB-deprojected maps (described in Section \ref{sec:null-tests}). We will refer to these masks as ``60\%'', ``40\%'' and ``CIB-depj'' from now on.

\subsection{Planck PR4 CMB lensing} \label{sec:data-Planck}

We combine the cross-correlation measurements using CMB lensing from ACT and \textit{Planck} at the likelihood level, properly accounting for their covariance as described in Section \ref{sec:covariance}. We use the PR4 \textit{Planck} lensing maps, which reconstruct the lensing convergence using CMB multipoles from $100<\ell<2048$ with the standard quadratic estimator in temperature and polarization \citep{Carron_2022}. Our analysis uses the reprocessed PR4 \texttt{NPIPE} maps that include around 8\% more data than the PR3 \textit{Planck} release from 2018. The resulting PR4 lensing maps increase the signal-to-noise ratio compared to the PR3 release by around 20\% due to pipeline improvements including optimal anisotropic filtering of the CMB fields. From this point onward, we use $\ell$ to denote lensing multipoles (before denoted by $L$), as we no longer refer to primary CMB multipoles in the remainder of this work.

\begin{figure*}
 \centering
 \includegraphics[width=0.75\linewidth]{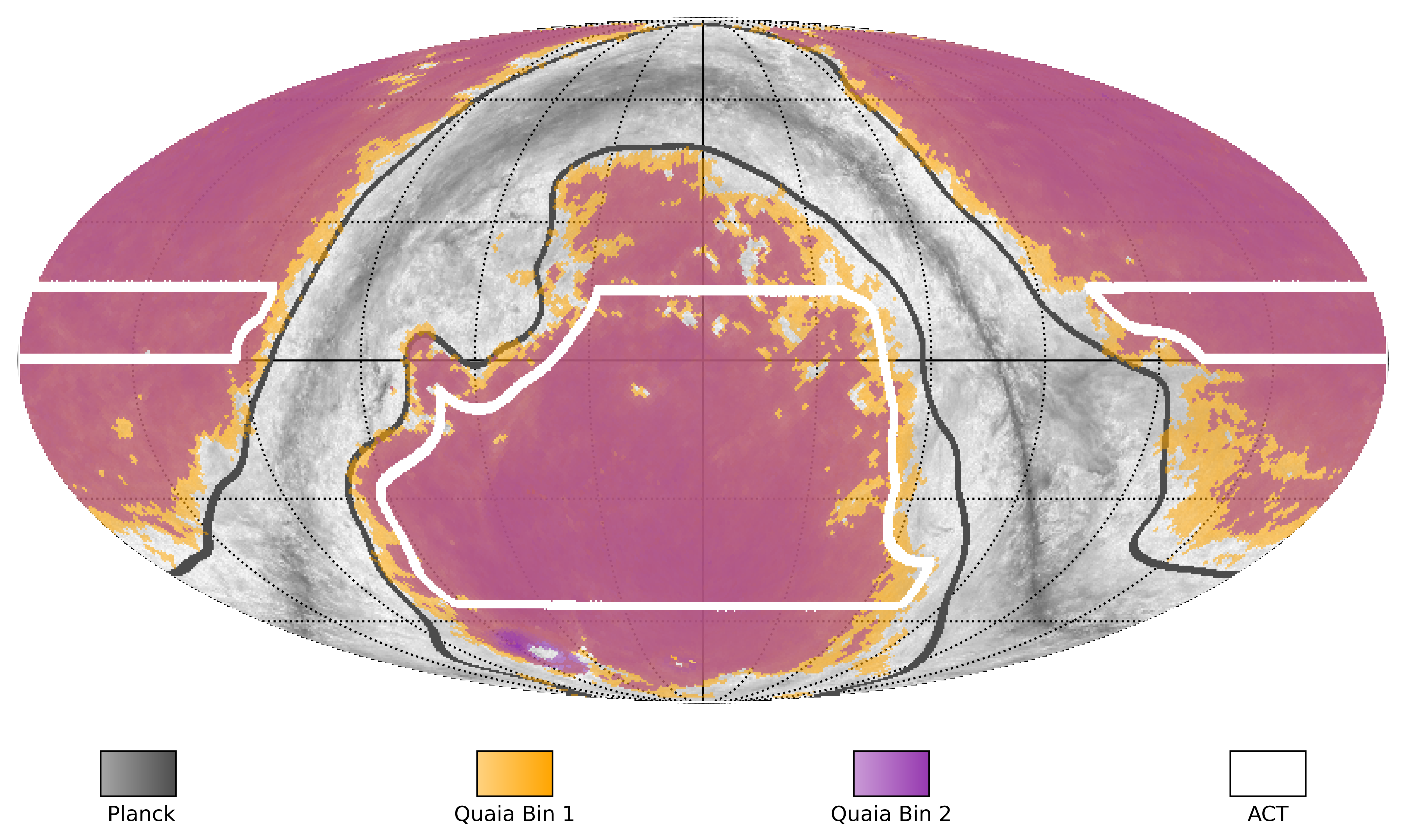}
 \caption{Overlap between the Quaia selection functions (orange for Bin 1, purple for Bin 2) and the CMB lensing maps from ACT DR6 (white contour) and \textit{Planck} PR4 (black contour), displayed on top of a Galactic dust map from \textit{Planck} \citep{Planck_dust} in grayscale. The overlap between the Quaia samples and the CMB lensing footprints is 98\% (Bin 1) and 99\% (Bin 2) with \textit{Planck}, and 34\% (Bin 1) and 38\% (Bin 2) with ACT.} 
 \label{fig:footprint}
\end{figure*}

Given the high overlap between the \textit{Planck} and Quaia footprints -- 98\% for Bin 1 and 99\% for Bin 2 -- we measure the auto-correlation of the quasar overdensity as well as its cross-correlation with \textit{Planck} lensing over the full Quaia footprint when performing the 2$\times$2pt analysis using \textit{Planck} lensing. Figure \ref{fig:footprint} shows the overlap of \textit{Planck} and ACT lensing footprints with the two Quaia samples.

\section{Theory}\label{sec:theory}

\subsection{Statistics of projected fields} \label{sec:theory-projected}

Both CMB lensing and the quasar overdensity are projected observables along the line of sight. Therefore, we need to compute the projected power spectra of the galaxy-galaxy and lensing-lensing auto-correlations as well as their cross-correlation in order to model our measurements. We use the Limber approximation \citep{Limber_1953,loverde&Afshordi2008} to compute
\begin{equation}
C_\ell^{gg} = \int d\chi \frac{W_g^2(z)}{\chi^2} P_{gg} \left(k = \frac{\ell + \frac{1}{2}}{\chi}, z\right) \; ,
\end{equation}
\begin{equation}
C_\ell^{\kappa g} = \int d\chi \frac{W_g(z) W_\kappa(z)}{\chi^2} P_{gm} \left(k = \frac{\ell + \frac{1}{2}}{\chi}, z\right) \; ,
\end{equation}
\begin{equation}
C_\ell^{\kappa\kappa} = \int d\chi \frac{W_\kappa^2(z)}{\chi^2} P_{mm} \left(k = \frac{\ell + \frac{1}{2}}{\chi}, z\right) \; ,
\end{equation}
where $\chi$ is the comoving distance to redshift $z$ along the line of sight, and $P_{mm}$, $P_{gg}$, and $P_{gm}$ are the three-dimensional power spectra of the matter, galaxy clustering and their cross-correlation, respectively. Here, $W_g$ and $W_\kappa$ are the galaxy and CMB lensing kernels defined by
\begin{align}
W_g(z) &= H(z) \frac{dN}{dz} \; , \\
W_\kappa(z) &= \frac{3}{2} \Omega_m H_0^2 (1 + z) \frac{\chi (\chi_{\star} - \chi)}{\chi_{\star}} \; .
\end{align}
Here, $dN/dz$ is the normalized redshift distribution of the quasars in our sample, described in Section \ref{sec:data-Quaia}, and $\chi_{\star}$ is the comoving distance to the surface of last scattering. We will restrict our analysis to linear scales with a $k_{\text{max}} = 0.15\,h\text{Mpc}^{-1}$ and use a linear bias model for the galaxy clustering and cross-correlation power spectra: $P_{gm} = b(z)P_{mm}$ and $P_{gg} = b^2(z)P_{mm}$. 

Past studies of the redshift evolution of quasar bias show that it evolves strongly with redshift; see, e.g.,~\citet{eBOSSqso,Alonsobias,Lin_2020} for spectroscopic quasars in SDSS and~\citet{dipompeo,Petter_2022} for photometrically-selected quasars. Therefore, we will assume the following parametric model from the eBOSS quasar sample \citep{eBOSSqso} for our bias evolution with redshift:
\begin{equation}
    b(z) = b^i_g \left[0.278\left((1+z)^2 - 6.565\right) + 2.393 \right] \; ,
    \label{eq:bias}
\end{equation}
where $b^i_g$ is a free parameter we fit for each redshift bin with $b^i_g=1$ recovering the best-fit found by eBOSS. Different models for the redshift dependence of the bias of Quaia quasars are explored in \cite{alonso2023}, finding no significant impact on cosmological parameters. 

Additionally, the angular power spectra considered in this analysis receive contributions from the lensing magnification bias, denoted by $\mu$; e.g., \citet{Narayan1989}. This effect depends on the number count slope $s$, which is the response of the galaxy number density to changes in magnitude, $s = d\log_{10} N / dm$, where $m$ is the apparent magnitude of the source. A detailed study of magnification bias in the Quaia sample was presented in \citet{alonso2023}, where the counts slope was found to be close to $s \approx 0.4$, and the resulting impact on the angular power spectra was shown to be negligible. To validate this for our specific analysis footprint, we re-evaluate the counts slope within the ACT DR6 footprint following \citet{alonso2023} and obtain values of $s_1 = 0.384 \pm 0.004$ and $s_2 = 0.419 \pm 0.004$ for redshift bins 1 and 2, respectively. Including magnification in our modeling, we estimate that its contribution modifies the power spectra by less than $0.02\sigma$ across all bandpowers and redshift bins used in our analysis. Therefore, since this effect is negligible within our error bars and does not significantly affect our results, we decide to neglect its contributions in our baseline analysis.

Our theoretical predictions are computed using \texttt{class\_sz}\footnote{\url{https://github.com/CLASS-SZ}} \citep{Bolliet_2023,bolliet2023classszioverview}, a Python and C-based code that extends the cosmological Boltzmann solver \texttt{CLASS} \citep{lesgourgues2011cosmiclinearanisotropysolving, Diego_Blas_2011} with machine learning acceleration. \texttt{class\_sz} includes a parallelized implementation of the Limber approximation and employs neural network emulators for the matter power spectrum based on \texttt{cosmopower} \citep{Spurio_Mancini_2022}. Although our analysis is restricted to only linear scales, we account for non-linear corrections to the matter power spectrum using the \texttt{HMCode} model \citep{10.1093/mnras/stw681, Mead:2020vgs} as implemented in \texttt{CLASS}.

\subsection{Redshift-evolution parametrization} \label{sec:theory-reconstruction}

Including the CMB lensing power spectrum, $C_\ell^{\kappa \kappa}$, in our analysis extends the traditional 2$\times$2pt framework to a 3$\times$2pt, to probe the growth of structure across a wider range of redshifts. In particular, this enables the measurement of the evolution of matter perturbations at redshifts beyond those directly traced by the Quaia quasar sample.

We model this evolution by introducing a redshift-dependent parametrization of the linear matter power spectrum that translates into an amplitude rescaling of $\sigma_8(z)$, defined as:
\begin{equation}
    \sigma^2_8(z) = \int_0^{\infty} dk \frac{k^2 P_{\text{lin}}(k,z)}{2\pi^2} |W_8(k)|^2 \; ,
\end{equation}
where $W_8(k)$ is the Fourier transform of a spherical-top-hat window function of radius $8 \,h^{-1}\text{Mpc}$, and $P_{\text{lin}}$ is the linear matter power spectrum.

Given the broad redshift kernels of the two Quaia samples and their overlap with the CMB lensing kernel, we scale the matter power spectrum with three free amplitude parameters $A_i$ \citep{DES_P_rescale} so that the first two ($A_1$ and $A_2$) are constrained mainly by the cross-correlation with the two Quaia samples while the last one ($A_3$), constrained by $C_\ell^{\kappa \kappa}$, encodes all the information beyond it. Therefore, the linear power spectrum is modified as 
\begin{equation}
P_{\text{lin}}^{\text{new}}(k,z) = P_{\text{lin}}(k,z) A(z) \; ,
\end{equation}
where
\begin{equation}
    A(z) = 
    \begin{cases} 
    A_1 & 0 \leq z < z_1 \; ,\\
    A_2 & z_1 \leq z < z_2 \; ,\\
    A_3 & z_2 \leq z \; , 
    \end{cases}
\label{eq:growth-rec}
\end{equation}
where $P_\mathrm{lin}(k,z)$ is the linear matter power spectrum computed with, e.g., a Boltzmann code like \texttt{CLASS} or \texttt{CAMB} \citep{Lewis:1999bs}, and $z_1=1.45$ and $z_2=3.0$ are chosen to minimize correlations between the new parameters of interest $\sigma_8 \sqrt{A_i}$ by approximately aligning with the redshift kernels of Quaia.

Although this modification applies only to the linear power spectrum (given the definition of $\sigma_8$) our full theory model includes non-linear corrections via \texttt{HMCode}. Therefore, following \cite{Farren_2024rla}, we account for this by modifying the non-linear power spectrum, $P_{\text{non-lin}}(k,z)$, as follows:
\begin{equation}
    \begin{aligned}
    P_{\text{non-lin}}^{\text{new}}(k,z) &= P_{\text{lin}}^{\text{new}}(k,z) + P_{\text{non-lin}}(k,z) - P_{\text{lin}}(k,z) \\
    &= P_{\text{lin}}(k,z) \left[ A(z) - 1 \right] + P_{\text{non-lin}}(k,z).
    \end{aligned}
\end{equation}
In this framework, $\sigma_8$ becomes degenerate with the amplitude parameters $A_i$, particularly since we restrict our 2$\times$2pt analysis to linear scales. To avoid this redundancy, we fix the amplitude of the primordial scalar perturbations, $A_s$, to its value from \textit{Planck} PR3 \citep{Planck2020}, $\ln(10^{10}A_s) = 3.041$,  thereby setting the normalization of the fiducial $\sigma_8(z)$. The free amplitude parameters $A_1$, $A_2$, and $A_3$ then quantify deviations from the $\Lambda$CDM history of structure growth at each of the redshift ranges. 

\section{Tomographic spectrum measurement} \label{sec:tomography-measurement}

\subsection{Bandpowers and analysis choices} \label{sec:measurement}

A first approach to estimating the angular power spectrum between two fields, $a$ and $b$, can be done by using the standard pseudo-$C_\ell$ estimator:
\begin{equation}
    \tilde{C}_{\ell}^{a b}=\frac{1}{2 {\ell}+1} \sum_{m=-{\ell}}^{\ell} a_{{\ell} m} b_{{\ell} m}^\ast
\end{equation}
where $a_{{\ell} m}$ and $b_{{\ell} m}$ are the spherical harmonic coefficients of $a$ and $b$.  This approach, however, is biased in the presence of masking (which is needed to account for survey geometry and Galactic foreground contamination) and induces a mode-coupling that needs to be accounted for. To do so we use the MASTER algorithm \citep{Hivon_2002} as implemented by the \texttt{NaMaster} code\footnote{\url{https://github.com/LSSTDESC/NaMaster}} \citep{Alonso_2019}. The MASTER method models the relation between the expectation value of the observed pseudo-$C_\ell$s and the true underlying angular power spectrum, $C_\ell^{ab}$, via a mode-coupling matrix $M_{\ell \ell'}^{ab}$:
\begin{equation}
    \left< \Tilde{C}_\ell^{ab} \right> = \sum_{\ell'} M_{\ell \ell'}^{ab} C_{\ell'}^{ab} + \Tilde{N}_\ell^{ab} \; ,
    \label{eq:pleudoCl}
\end{equation}
where $\Tilde{N}_\ell^{ab}$ is the noise pseudo-$C_\ell$, which is only non-zero in the auto-correlation case ($a=b$). The MASTER algorithm approximately inverts this relation by assuming the power spectrum to be piecewise constant within the $\ell$ range of each bandpower bin.

To perform the 2$\times$2pt measurement, we compute the angular power spectra of the quasar overdensity auto-correlations and their cross-correlation with CMB lensing -- denoted $C_\ell^{gg}$ and $C_\ell^{\kappa g}$, respectively -- for each of the redshift samples. We do not include the cross-bin quasar correlations, as their signal-to-noise ratio is low. In addition, they are also more susceptible to systematic effects that may be common between the different redshift bins.

When performing CMB lensing reconstruction using a quadratic estimator on a masked sky, the signal is convolved with the mask in a more complex way than captured by the standard MASTER formalism \citep{qu2023atacama}. To account for this effect, we take the mask in the CMB lensing field to be the square of the original CMB mask, which has been proven to be a good approximation \citep[see, e.g.,][]{Farren_2024}. We additionally apply a multiplicative transfer function, described in Section~\ref{sec:tf}, to correct for residual bias. This leads to a power spectrum recovery within 3.5\% as discussed in Section \ref{sec:simulations}.

For both the quasar overdensity field and CMB lensing convergence we use \texttt{HEALPix} maps of \texttt{nside} = 512. We bin all power spectra with a linear $\ell$ separation of width $\Delta \ell = 30$ between $\ell = 2$ and $\ell = 3\,{\texttt{nside}}-1 = 1535$ to avoid bias from the pseudo-$C_\ell$ method. We discard the first bandpower ($\ell<32$) in all power spectra, where systematic contamination was found to be most relevant \citep{alonso2023}, and to mitigate potential mis-estimation of the lensing-reconstruction mean-field on those scales \citep{qu2023atacama, Madhavacheril_2024}. On small scales, we impose a scale cut of $k_{\text{max}} = 0.15\,h\text{Mpc}^{-1}$ to ensure the validity of the linear bias model adopted in our theoretical predictions. This choice is more conservative than the one used in \citet{alonso2023} and, when combined with a marginalization over the shot noise amplitude, leads to broader posterior constraints compared to this earlier analysis. A detailed discussion of these and other changes to the analysis can be found in Appendix \ref{app:Planck-reanalysis}. 

\begin{figure*}
 \centering
 \includegraphics[width=\linewidth]{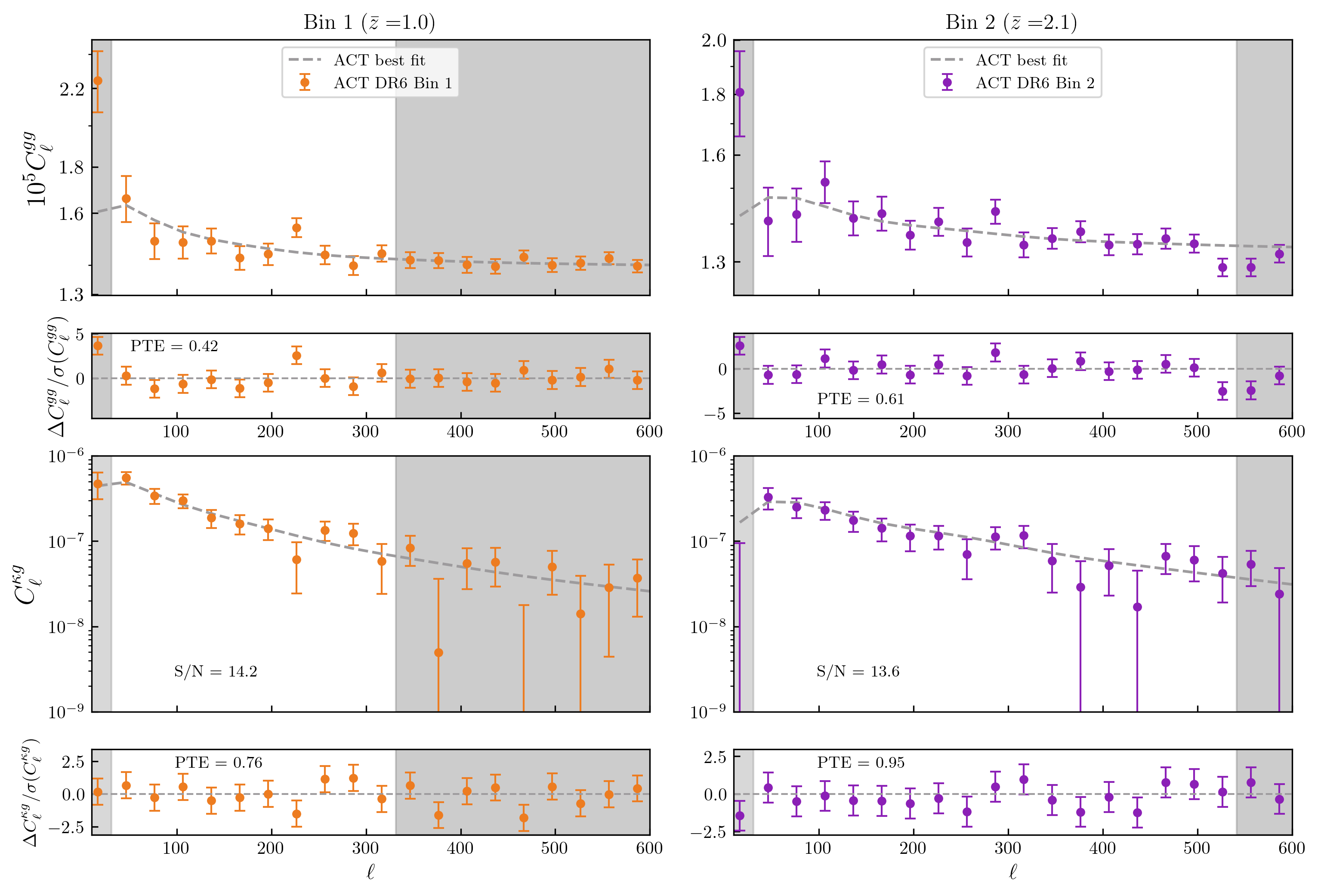}
 \caption{Measurements of $C_\ell^{gg}$ on the ACT footprint (top) and $C_\ell^{\kappa g}$ (bottom) for each of the two redshift bins considered. Dashed gray lines indicate the best-fit model from the joint fit to both redshift bins and gray bands indicate the excluded bandpowers in our analysis. The cross-correlation is detected within the analysis range with SNRs of 14.2 and 13.6 for Bin 1 and Bin 2, respectively, for a total detection of $19.6\sigma$. Panels below each measurement show the model residuals. The joint fit to $C_\ell^{\kappa g}$ and $C_\ell^{gg}$ across both bins gives a total $\chi^2$ of 41.8 for 54 bandpowers and 7 fitted parameters (see Section~\ref{sec:cosmoanal-likelihood}). We estimate a PTE of 0.69.}
 \label{fig:measurement}
\end{figure*}

This scale cut translates into redshift bin-dependent cuts in multipole space: $\ell_{\text{max}} = k_{\text{max}}\chi(\bar{z})$, where $\chi(\bar{z})$ is the comoving distance to the mean redshift of each bin $\bar{z}$, which we compute by assuming the same fiducial cosmology we use to generate our simulations. This yields $\ell_{\text{max}} = 316$ and $526$ for Bin 1 and Bin 2, respectively. The resulting auto- and cross-power spectra are shown in Figure~\ref{fig:measurement}.

In the 3$\times$2pt case, we include the CMB lensing auto-spectrum $C_{\ell}^{\kappa \kappa}$, using the pre-computed and publicly available bandpowers from the ACT DR6 lensing analysis \citep{Madhavacheril_2024, qu2023atacama}. We appropriately account for their covariance with the 2$\times$2pt block as described in Section \ref{sec:covariance}.

\subsection{Simulations} \label{sec:simulations}

To compute the multiplicative transfer functions to be described in Section \ref{sec:tf}, build the covariance matrices, and validate our pipeline, we make use of Gaussian simulations of both the CMB lensing convergence and the quasar overdensity fields, ensuring that the two are appropriately correlated.

\textit{ACT simulations --} We use 400 Gaussian CMB lensing simulations publicly available as part of the ACT DR6 products. These simulations were constructed by displacing a randomly drawn CMB realisation with a Gaussian lensing convergence field, then adding realistic survey noise \citep{Atkins_2023}, before masking and passing the resulting lensed CMB maps through the same reconstruction pipeline that is applied to the data. Further information about these simulations can be found in \cite{qu2023atacama}.

To generate correlated simulations of the quasar overdensity field, we begin by constructing fiducial power spectra $C_{\ell \text{,fid}}^{gg}$ and $C_{\ell \text{,fid}}^{\kappa g}$. These are generated by fitting the observed spectra with fixed cosmological parameters and determining the best-fit values for the galaxy bias and shot noise parameters\footnote{To perform this fit, we use a Gaussian covariance matrix computed with \texttt{NaMaster} \citep{Alonso_2019,García-García_2019}. After generating the simulations and constructing the final covariance matrix, we repeat the fit and confirm that changes in the $C_{\ell,\text{fid}}$ are negligible (they stay below 5\% across the analysis $\ell$ range).}. Then, following \citet{Farren_2024}, we generate the spherical-harmonic coefficients of the quasar field $a_{\ell m}^g$ by decomposing them into three contributions: a component correlated with CMB lensing, an uncorrelated component, and the shot-noise term. Each realization is then generated as follows:
\begin{equation}
    a_{\ell m}^g=\frac{C_{\ell \text {, fid }}^{\kappa g}}{C_{\ell, \text { fid }}^{\kappa \kappa}} a_{\ell m}^\kappa+a_{\ell m}^{g, \text { uncorr.}}+a_{\ell m}^{g, \text { noise }} \; ,
\end{equation}
where $a_{\ell m}^\kappa$ are the spherical-harmonic coefficients of the $\kappa$ field used as input to the corresponding CMB lensing simulation. This construction ensures that the quasar field is correlated with the lensing field according to the fiducial cross-spectrum $C_{l\text{,fid}}^{\kappa g}$. The second term, representing the uncorrelated component, is generated as a Gaussian random field with variance
\begin{equation}
    \left\langle a_{\ell m}^{g, \text {uncorr.}}\left(a_{\ell^{\prime} m^{\prime}}^{g, \text {uncorr.}}\right)^*\right\rangle=\delta_{\ell \ell^{\prime}} \delta_{m m^{\prime}}\left(C_{\ell, \text {fid}}^{g g}-\frac{\left(C_{\ell, \text {fid}}^{\kappa g}\right)^2}{C_{\ell, \text {fid}}^{\kappa \kappa}}\right) \; .
\end{equation}
This ensures that the auto-spectra of the quasar simulations follow the fiducial auto-spectrum $C_{\ell, \text {fid}}^{g g}$. We note that these simulations do not account for correlations between different redshift bins. As will be discussed in Section \ref{sec:covariance}, these contributions to the covariance matrix are computed analytically. Finally, the shot-noise term $a_{\ell m}^{g, \text { noise }}$ is drawn independently to reproduce the expected shot-noise level. 

With these simulated maps, we compute the auto- and cross-power spectra in the same way as we do for the data. Since we expect their mean to match the input fiducial spectra by construction, we can validate our pipeline by comparing the mean spectra measured from the simulations $\left< C_{\ell, \text{sim}}^{ab}\right>$ to the corresponding fiducial spectra $ C_{\ell,\text{fid}}^{ab} $.

For this comparison, we convolve the fiducial spectra with the appropriate (unbinned) mode-coupling matrix, bin them, and then deconvolve them accounting for the pixel window function. This follows the same procedure used to treat the theoretical prediction in our likelihood. For the cross-spectra, we also apply the multiplicative transfer function, described in Section~\ref{sec:tf}.

\begin{figure*}
 \centering
 \includegraphics[width=\linewidth]{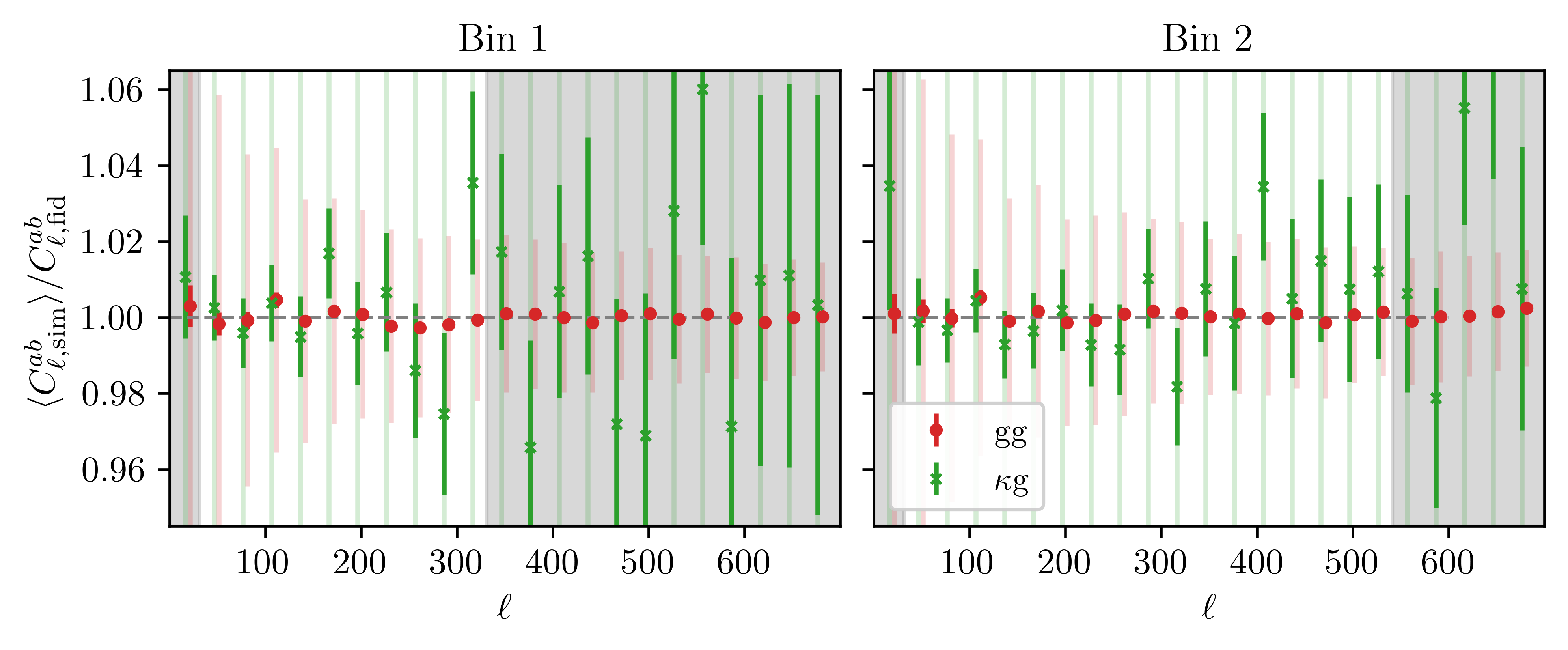}
 \caption{Recovery of $C_\ell^{\kappa g}$ (green) and $C_\ell^{g g}$ (red) from simulations. Lighter error bars represent the statistical uncertainty in the measurement, while darker bars indicate the error on the mean over 400 simulations. Gray bands indicate the excluded bandpowers in our analysis. Residuals are within 3.5\% for the cross-spectra and sub-percent for the auto-spectra across the analysis range.}
 \label{Fig:Sim_residuals}
\end{figure*}

We find good agreement between the recovered and input spectra as shown in Figure~\ref{Fig:Sim_residuals}. The mean cross-spectra recovery is accurate to within $3.5\%$ (limited by Monte Carlo error from the 400 simulations), and the auto-spectra to within $0.5\%$ over the analysis range. These are both well within the measurement error, corresponding to $0.1\sigma$ and $0.09\sigma$ deviations, respectively.

\textit{Planck simulations --} We perform a reanalysis of the cross-correlation using \textit{Planck} lensing in Appendix \ref{app:Planck-reanalysis}, and present the joint results with ACT in Section \ref{sec:results-2x2pt-Planck}. To compute the corresponding covariance matrix and transfer functions, we follow the same approach as for the ACT measurement. We use 480 FFP10 CMB simulations presented in \cite{Carron_2022}. These are constructed in a similar way as the ACT convergence simulations and also include lensing reconstructions computed using the \textit{Planck} PR4 lensing pipeline. Correlated quasar overdensity simulations are constructed following the same procedure outlined above. We also assess the recovery of the power spectra and find agreement within 2.7\% for the cross-correlations and 0.5\% for the auto-correlations over the analysis range.

\textit{Joint simulations --} As we will show in Section~\ref{sec:results}, we find good agreement between cosmological parameters inferred using the \textit{Planck} and ACT measurements and so combine them to derive joint constraints. To account for the covariance between the two correctly, we use ``ACT-like'' reconstructions of the \textit{Planck} input simulations described above. As detailed in \cite{qu2023atacama}, these simulations are constructed by processing the FFP10 \textit{Planck} simulations through the ACT DR6 pipeline, which uses different CMB scales and sky area. As a result, these simulations capture correlations between the two lensing reconstructions arising from shared CMB fluctuations, but do not include instrumental and atmospheric noise, which is uncorrelated between the surveys.

\subsection{Transfer function}\label{sec:tf}

CMB lensing convergence maps from both ACT and \textit{Planck} are obtained via the lensing reconstruction of masked CMB maps. This masking introduces a well-known bias in the reconstructed lensing maps, resulting in a misnormalization of the lensing spectra \citep[e.g.,][]{Benoit_Levy_2013, Carron2023}.

This effect is footprint-dependent so it must be re-evaluated on a case-by-case basis. Previous cross-correlation analyses have accounted for this effect by computing a Monte Carlo (MC) transfer function from Gaussian simulations \citep[e.g.,][]{Qu_DESI_2025, Farren_2024}. In our case, we compute a separate transfer function for each redshift bin, as differences in the selection functions lead to slightly different sky masks.

For a cross-correlation this MC correction can be defined as
\begin{equation}
    A^{\text{MC}}_\ell = \frac{\left<\kappa_{\text{in, $\kappa$-mask}} , \kappa_{\text{in, $g$-mask}}\right>}{\left<\hat{\kappa}, \kappa_{\text{in, $g$-mask}}\right>} \; ,
\end{equation}
where the angular cross-spectrum between two fields $X$ and $Y$ is denoted as $C_\ell^{XY}=\left<X,Y\right>$, $\kappa_{\text{in}}$ is the input lensing convergence map and $\hat{\kappa}$ is the corresponding masked CMB lensing reconstruction. Subscripts $\kappa$-mask and $g$-mask refer to the CMB-lensing and galaxy masks, respectively, applied to the lensing convergence input.

\begin{figure}
 \centering
 \includegraphics[width=\linewidth]{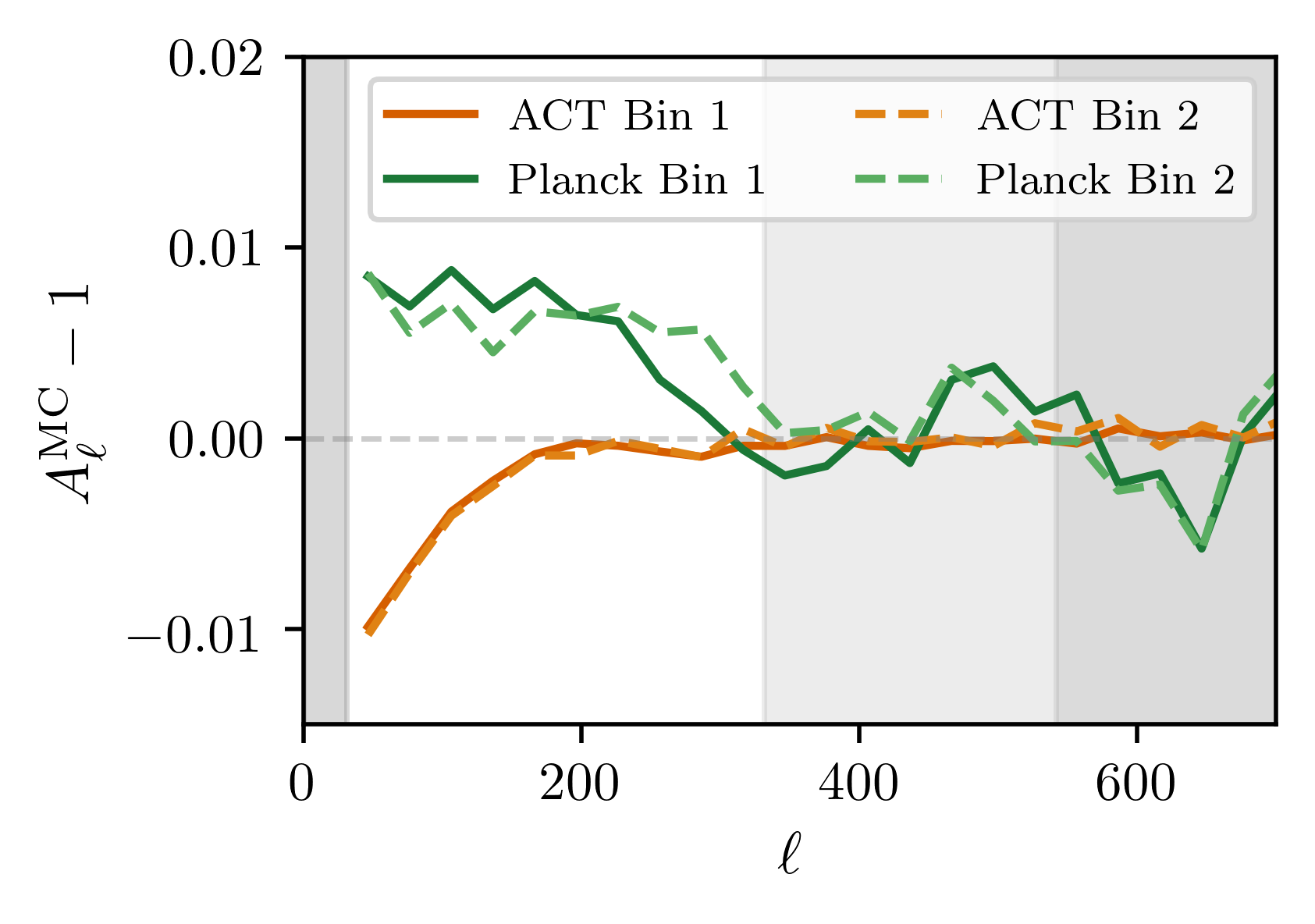}
 \caption{Monte Carlo normalization corrections for the cross-correlations of ACT DR6 and \textit{Planck} PR4 lensing with the two Quaia redshift bins. The dark gray band denotes the multipole range excluded for Bin 2, while the lighter gray band indicates additional scales excluded for Bin 1. The shaded region on the left corresponds to the first multipole bin, which is excluded in both cases. Corrections remain below 1\% across all scales used in the analysis.}
 \label{Fig:transfer_function}
\end{figure}

To incorporate this correction, we bin $A^{\text{MC}}_\ell$ using the same scheme applied to our power spectra assuming the correction is piecewise constant across each $\ell$ bin and apply it directly to both the measured and simulated cross-correlations:
\begin{equation}
    C_{\ell}^{\hat{\kappa} g} \rightarrow A^{\text{MC}}_\ell C_{\ell}^{\hat{\kappa} g} \; .
\end{equation}
We show the resulting Monte Carlo transfer functions for the two Quaia redshift bins in cross-correlation with ACT DR6 and \textit{Planck} PR4 lensing in Figure~\ref{Fig:transfer_function}. Across the full analysis range, the correction remains below 1\%.

\subsection{Covariance matrices} \label{sec:covariance}

\begin{figure}
 \centering
 \includegraphics[width=\linewidth]{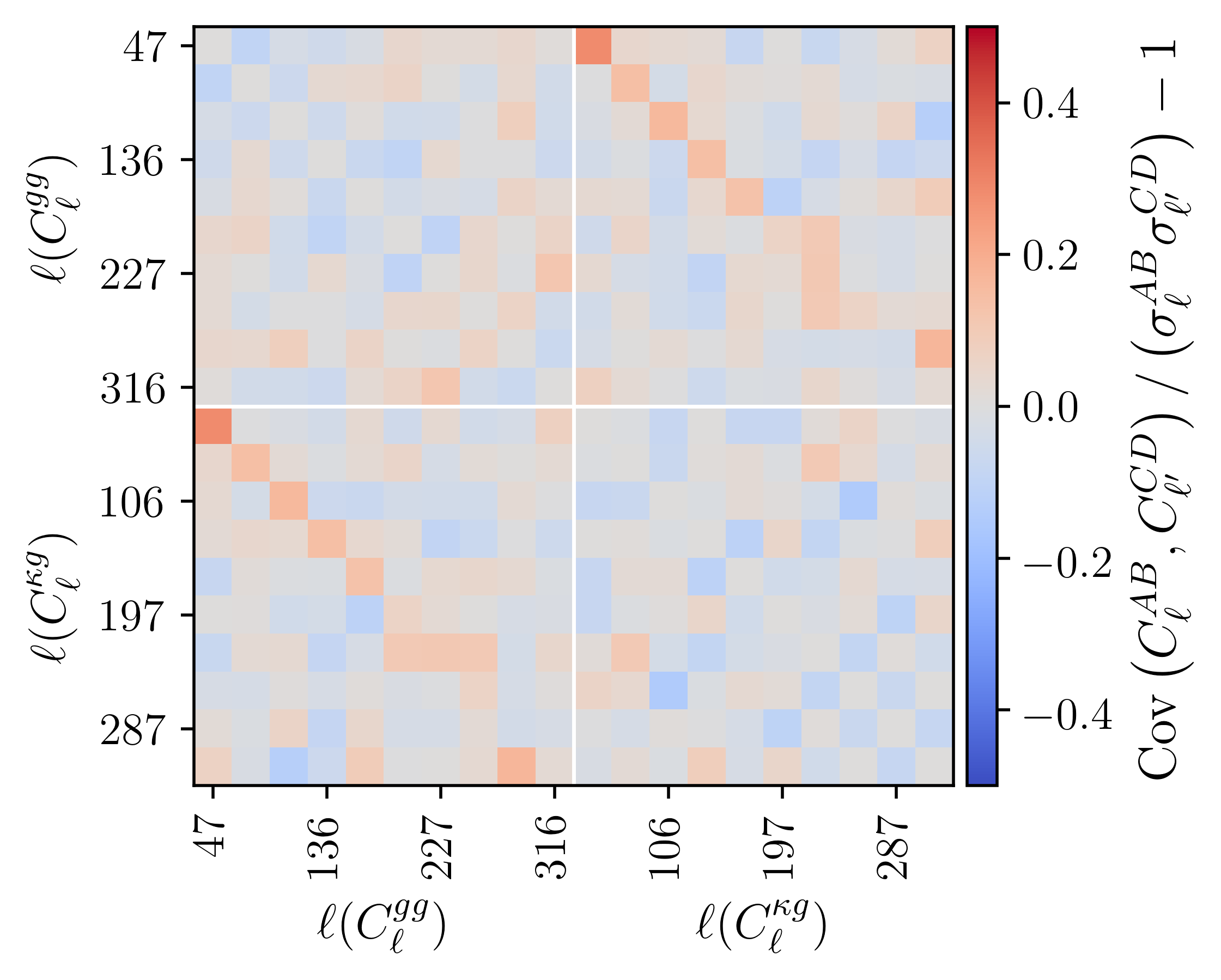}
 \caption{Correlation matrix for the 2$\times$2pt measurement for Bin~1 and ACT DR6 lensing. Here, $\text{Cov}(C^{AB}_{\ell},C^{CD}_{\ell'})$ is the covariance between bandpowers $\ell$ and $\ell'$ of spectra $C^{AB}_\ell$ and $C^{CD}_\ell$, and $\sigma_\ell^{AB}$ and  $\sigma_{\ell'}^{CD}$ are their respective standard deviations. The diagonal blocks have been nulled for visualization.}
 \label{fig:cov_0_ACT}
\end{figure}

\textit{2$\times$2pt covariance --} We construct the covariance matrix for the data vector $\left[ {C_\ell^{g_1 g_1}, C_\ell^{\kappa g_1}, C_\ell^{g_2 g_2}, C_\ell^{\kappa g_2}}\right]$ using a combination of simulations and analytic calculations. The diagonal blocks of the covariance matrix, which do not mix different redshift bins, are estimated using the 400 Gaussian simulations described in Section~\ref{sec:simulations}. These capture both multipole correlations and the covariance between $C_\ell^{g_i g_i}$ and $C_\ell^{\kappa g_i}$ within each redshift bin. An example block for the cross-correlation measurement in redshift Bin 1 is shown in Figure~\ref{fig:cov_0_ACT}.

Since our simulations do not include correlations between different redshift bins, we compute the off-diagonal blocks between redshift bins, such as $\mathrm{Cov}(C_\ell^{g_1 g_1}, C_{\ell'}^{g_2 g_2})$, analytically, using the Gaussian covariance module from \texttt{NaMaster} \citep{Alonso_2019,García-García_2019}. This calculation requires the following fiducial spectra: the cross-spectra of quasar overdensities between different redshift bins $C_\ell^{g_i g_j}$, their cross-correlations with CMB lensing $C_\ell^{\kappa g_i}$, and the lensing convergence auto-spectrum $C_\ell^{\kappa \kappa}$. The first two are computed using the fiducial cosmology and bias parameters used to generate the simulations (see Section~\ref{sec:simulations}), while $C_\ell^{\kappa \kappa}$ is measured from the simulations, including reconstruction noise.

We find a correlation between $C_{\ell}^{g_i g_i}$ and $C_{\ell}^{\kappa g_i}$ at the same $\ell$ bin of up to 30\% for Bin~1 and 35\% for Bin~2. In contrast, correlations between different redshift bins -- for both $C_\ell^{g g}$ and $C_\ell^{\kappa g}$ -- are found to be almost negligible (less than 1\%).

When inverting the covariance matrix in our likelihood analysis, we correct for the bias introduced by estimating the inverse from a finite number of simulations using the Hartlap correction factor \citep{Hartlap_2006}:
\begin{equation}
    \alpha_{\text{cov}} = \frac{N_{\text{sims}} - N_{\ell\text{ bins}} - 2}{N_{\text{sims}} -1 .} \; .
\end{equation}
With 400 simulations and 54 $\ell$-bins, this yields a Hartlap factor of approximately $\alpha_{\text{cov}} = 0.86$.

For the measurement using \textit{Planck} CMB lensing, we follow the same approach: the diagonal blocks (i.e., not mixing redshift bins) of the covariance matrix are estimated using the simulations described in Section~\ref{sec:simulations}, while the off-diagonal contributions between redshift bins are computed analytically. The fiducial spectra $C_{\ell}^{g_i g_j}$ and $C_{\ell}^{\kappa g_i}$ are computed using the same fiducial cosmology and bias parameters used to generate the quasar simulations correlated with \textit{Planck}. The $C_\ell^{\kappa \kappa}$ spectrum is measured from simulations and includes the effect of the \textit{Planck} PR4 reconstruction noise.

Since we find good agreement in cosmological parameters between the two datasets (see Section \ref{sec:results}), we combine the ACT and \textit{Planck} measurements using the \textit{Joint} simulations described in Section \ref{sec:simulations}, which capture correlations between ACT and \textit{Planck} measurements within the same redshift bin. As we also include quasar auto-correlation measurements in both the ACT and \textit{Planck} footprints, we evaluate their correlation by measuring the auto-spectra of the same quasar overdensity simulations in each footprint. As before, the off-diagonal contributions between different redshift bins are computed analytically.

As expected, we find a strong correlation -- up to 75\% -- between the quasar auto-spectra in the two footprints for the same redshift bin. The correlations between ACT and \textit{Planck} lensing cross-spectra are found to be up to 48\% and 45\% in Bins~1 and 2, respectively, for the same $\ell$-bins, due to the identical quasar sample used in both cases.

\textit{3$\times$2pt covariance --} In Section~\ref{sec:results-3x2pt}, we present the results extending our analysis by including the ACT DR6 CMB lensing auto-correlation, $C_\ell^{\kappa \kappa}$, in combination with the previously described 2$\times$2pt measurements. To model the covariance for this joint 3$\times$2pt data vector, we estimate the correlations between $C_\ell^{\kappa \kappa}$ and all other observables using the same set of ACT lensing convergence simulations described above.

Specifically, we use the bias-subtracted CMB lensing auto-power spectra measured from the ACT simulations to compute the cross-covariance with the quasar auto- and cross-correlations. These simulations naturally account for correlations with measurements using both ACT and \textit{Planck} CMB lensing.

We find correlations between the ACT lensing auto-spectrum and its cross-correlation with quasars to range between 25--35\%, while correlations with the \textit{Planck} cross-correlation are at most 15\%. Additionally, correlations between the quasar and lensing auto-spectra remain below $\sim 20\%$ for both redshift bins and footprints.

Finally, we note that the off-diagonal elements of the simulation-based covariance between multipole bins are noisy, introducing spurious correlations across multipoles. If unaccounted for, this can lead to an artificial variance cancellation and overly tight parameter constraints. To mitigate this, we adopt a more conservative approach and keep only the diagonal and first off-diagonal $\ell$ elements of the covariance between $C_\ell^{\kappa \kappa}$ and the 2$\times$2pt block. This choice does not shift the inferred mean parameter values but results in a mild broadening of the posteriors by around $6.5\%$.

\section{Systematics and contamination tests} \label{sec:tests}

To ensure the robustness of our measurements, we conduct a suite of tests aimed at identifying potential sources of contamination in the ACT DR6 CMB lensing reconstruction maps. These tests assess whether systematics could bias our results by inducing additional correlations with the quasar overdensity field. For an exhaustive analysis of potential contaminants affecting the Quaia sample see \cite{alonso2023} and \cite{Storey_Fisher_2024}.

Here, we focus on two complementary classes of systematics tests: bandpower null tests (Section \ref{sec:null-tests}), and a simulation-based test designed to assess extragalactic foreground contamination (Section \ref{sec:tests-simbased}).

\subsection{Bandpower null tests} \label{sec:null-tests}

The CMB lensing reconstruction can be affected by various astrophysical foregrounds, including the tSZ effect -- caused by inverse Compton scattering of CMB photons by hot electrons in galaxy clusters~\citep{1970SZ} -- and the CIB, originating from dusty, star-forming galaxies. These components can contaminate the temperature maps used in the reconstruction and, if correlated with the quasar overdensity field, bias the cross-correlation measurement \citep{Osborne_2014,van_Engelen_2014}.

To test for such contamination, we perform null tests using differences between cross-correlations involving different versions of the reconstructed convergence maps. We compute $C_\ell^{\kappa_1 g} - C_\ell^{\kappa_2 g}$, where $\kappa_1$ and $\kappa_2$ denote different lensing reconstructions. A statistically significant non-zero result could indicate the presence of systematics or contamination in one or both of the maps.

We use the following convergence maps\footnote{Publicly available at \url{https://lambda.gsfc.nasa.gov/product/act/actadv_dr6_lensing_maps_info.html}.}:

\begin{itemize}
    \item Reconstruction using polarization-only data (MVPOL), expected to be largely free of extragalactic contamination.
    \item Temperature-only reconstruction (TT), more susceptible to foregrounds such as tSZ and CIB.
    \item Minimum-variance reconstruction (MV) combining temperature and polarization. This is the baseline map used in our analysis.
    \item Minimum-variance map with an explicit deprojection \citep{Remazeilles_2011_ILC} of the CIB component (CIB-deprojected MV) which includes information from \textit{Planck} $353\,\text{GHz}$ and $545\,\text{GHz}$.
\end{itemize}

Because significant extragalactic contamination is only expected in the temperature data, null tests involving the MVPOL map can effectively isolate such systematics. For example, a non-zero difference between MVPOL and TT maps would suggest contamination in the temperature-only reconstruction maps. 

We further investigate extragalactic, frequency-dependent contamination by comparing the MV and TT convergence maps at different frequencies, namely $150\,\text{GHz}$ and $90\,\text{GHz}$ (denoted by ``f150'' and ``f090'' prefixes, respectively). 

All cross-correlations are computed following the same procedure used in our main analysis. In addition, we estimate the covariances for each null test using cross-correlations between the 400 quasar simulations and the corresponding lensing reconstructions for each version of the convergence map.

We consider a test to be consistent with the null hypothesis (and, therefore, ``passing") if its probability-to-exceed (PTE) is above 0.05. We find no low failures under this condition. Table~\ref{table:nulls-2} summarizes the results for the two redshift bins.

We note that some of our tests have low $\chi^2$ with PTE values above 0.95, which may suggest an overestimation of the covariance. To investigate this further, we compute the cross-correlation between the quasar overdensity and the curl-mode lensing map ($C_\ell^{\omega g}$). Since the curl mode is expected to contain no cosmological signal, this essentially tests our covariance estimation. We find no evidence for the overestimation of our covariance, as neither of the PTE values of the curl tests appear to be high.

\begin{figure}
 \centering
 \includegraphics[width=\linewidth]{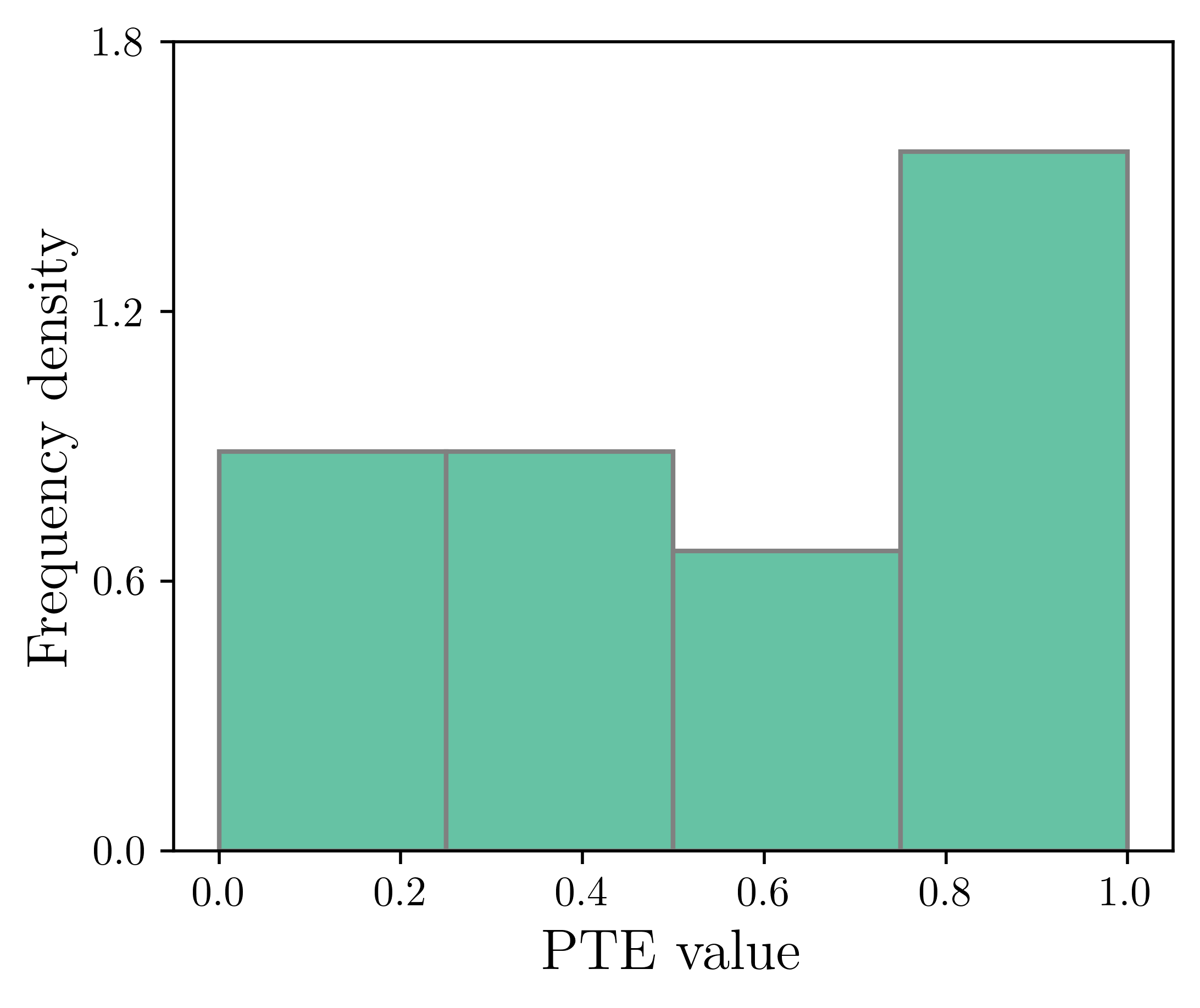}
 \caption{Histogram of the PTE values of the total 18 null tests performed across both redshift bins. The distribution is consistent with uniform, as confirmed by a Kolmogorov–Smirnov test (PTE = 0.43).}
 \label{fig:PTEhist}
\end{figure}

To assess further these high ``failures'', we check that the frequency of PTE values is consistent with a uniform distribution\footnote{We note that some of these tests are correlated, so technically the PTE distribution should only be approximately uniform.}. To do so, we compile all 18 PTE values (see Figure~\ref{fig:PTEhist}) and perform a Kolmogorov--Smirnov test which yields a PTE of 0.43. This indicates that the distribution of our tests are compatible with a uniform distribution, and that these high PTE values are compatible with statistical fluctuations.

Finally, we test for spatial inhomogeneities in the quasar sample by comparing results derived using the 60\% and 40\% Galactic masks described in Section~\ref{sec:data}. We repeat the MV$–$MVPOL test within the more restricted 40\% footprint and compare MV reconstructions between the two masks. We find no evidence of significant spatial inhomogeneities near the Galaxy in either redshift bin. 

\begin{table}
\centering
\begin{tabular}{lcc}
\toprule
\multicolumn{1}{c}{} & \multicolumn{2}{c}{\textbf{PTE values}} \\
\textbf{\textbf{Null test}} & Bin 1 & Bin 2 \\
\midrule
MV$-$MVPOL & 0.864 & 0.931 \\
MV$-$TT & 0.138 & 0.978 \\
TT$-$MVPOL & 0.601 & 0.974 \\
MV$-$CIB deprojected MV (CIB-depj mask) & 0.348 & 0.513 \\
\midrule
f150\_MV$-$f090\_MV & 0.208 & 0.702 \\
f150\_TT$-$f090\_TT & 0.491 & 0.909 \\
\midrule
MV (40\% mask)$-$MVPOL (40\% mask) & 0.301 & 0.929 \\
MV$-$MV (40\% mask) & 0.455 & 0.804 \\
\midrule
Curl & 0.166 & 0.208 \\
\bottomrule
\end{tabular}
\caption{Summary of null tests performed in the two redshift samples described in Section \ref{sec:null-tests}. For each test $C_\ell^{\kappa_1 g} - C_\ell^{\kappa_2 g}$ we specify the convergence maps $\kappa_1, \kappa_2$ used. The final row corresponds to the cross-correlation of the quasar overdensity with the lensing curl mode $C_\ell^{\omega g}$. We perform a total of 18 null tests, nine in each redshift bin.}
\label{table:nulls-2}
\end{table}

\subsection{Simulation-based tests for extragalactic foregrounds} \label{sec:tests-simbased}

To assess potential contamination from extragalactic foregrounds, we perform dedicated tests using realistic foreground maps from the \texttt{WebSky} simulations \citep{Stein_2020}. Following the approach in \cite{maccrann2023atacama}, we assume that such foregrounds only affect the CMB temperature, not polarization -- i.e., $T = T_{\text{CMB}} + T_{\text{fg}}$ -- since signals such as the tSZ effect and the CIB are not expected to be significantly polarized at our current sensitivity levels. Under the assumption that foregrounds are uncorrelated with the primary CMB temperature, the induced bias in the cross-correlation of the temperature-only lensing reconstruction with quasars can be expressed as
\begin{equation}
    \Delta C_{\ell}^{\kappa g}=\left\langle\mathcal{Q}\left(T_{\text{fg}}, T_{\text{fg}}\right) g\right\rangle \; ,
\end{equation}
where $\mathcal{Q}(T_X, T_Y)$ denotes the lensing reconstruction quadratic estimator applied to two fields $T_X$ and $T_Y$, and $g$ is the quasar overdensity. The angular cross-spectrum between any two fields $A$ and $B$ is denoted as $C_\ell^{AB} = \langle A  B \rangle$.

To evaluate this potential bias, we measure the cross-correlation between the lensing signal reconstructed from the foreground-only temperature maps presented in \cite{maccrann2023atacama} and a mock quasar overdensity field derived from the \texttt{WebSky} simulation. This mock catalog is constructed by populating the \texttt{WebSky} halo catalog with galaxies according to a simple halo occupation distribution (HOD) that matches the Quaia sample, detailed in Appendix~\ref{app:HOD}. The resulting galaxy catalog is then sampled to reproduce both the redshift distribution and angular number density of the observed quasars. We then validate these mock catalogs by comparing their angular power spectra to our measurements (see Appendix \ref{app:Websky}).

To quantify the significance of the contamination relative to our constraints, we compute the ratio of the induced bias to the statistical uncertainty on the cross-correlation amplitude $A_{\times}$:
\begin{equation}
    \frac{\Delta A_{\times}}{\sigma\left(A_{\times}\right)}=\frac{\sum_{\ell \ell^{\prime}} \Delta C_{\ell}^{\kappa g} \mathbb{C}^{-1}_{\ell \ell^{\prime}} C_{\ell^{\prime}, \mathrm{fid}}^{\kappa g}}{\left[\sum_{\ell \ell^{\prime}}{C }_{\ell, \mathrm{fid}}^{\kappa g} \mathbb{C}^{-1}_{\ell \ell^{\prime}} C_{\ell^{\prime}, \mathrm{fid}}^{\kappa g}\right]^{1/2}} \;,
\end{equation}
where $\Delta C_{\ell}^{\kappa g}$ is the measured cross-correlation between the mock quasar catalog and the lensing map reconstructed from foreground-only temperature maps, $C_{\ell}^{\kappa g, \mathrm{fid}}$ is the fiducial theoretical cross-spectrum (see Section~\ref{sec:simulations}), and $\mathbb{C}^{-1}_{\ell \ell'}$ is the inverse of the covariance matrix used in our main analysis.

\begin{figure*}
 \centering
 \includegraphics[width=\linewidth]{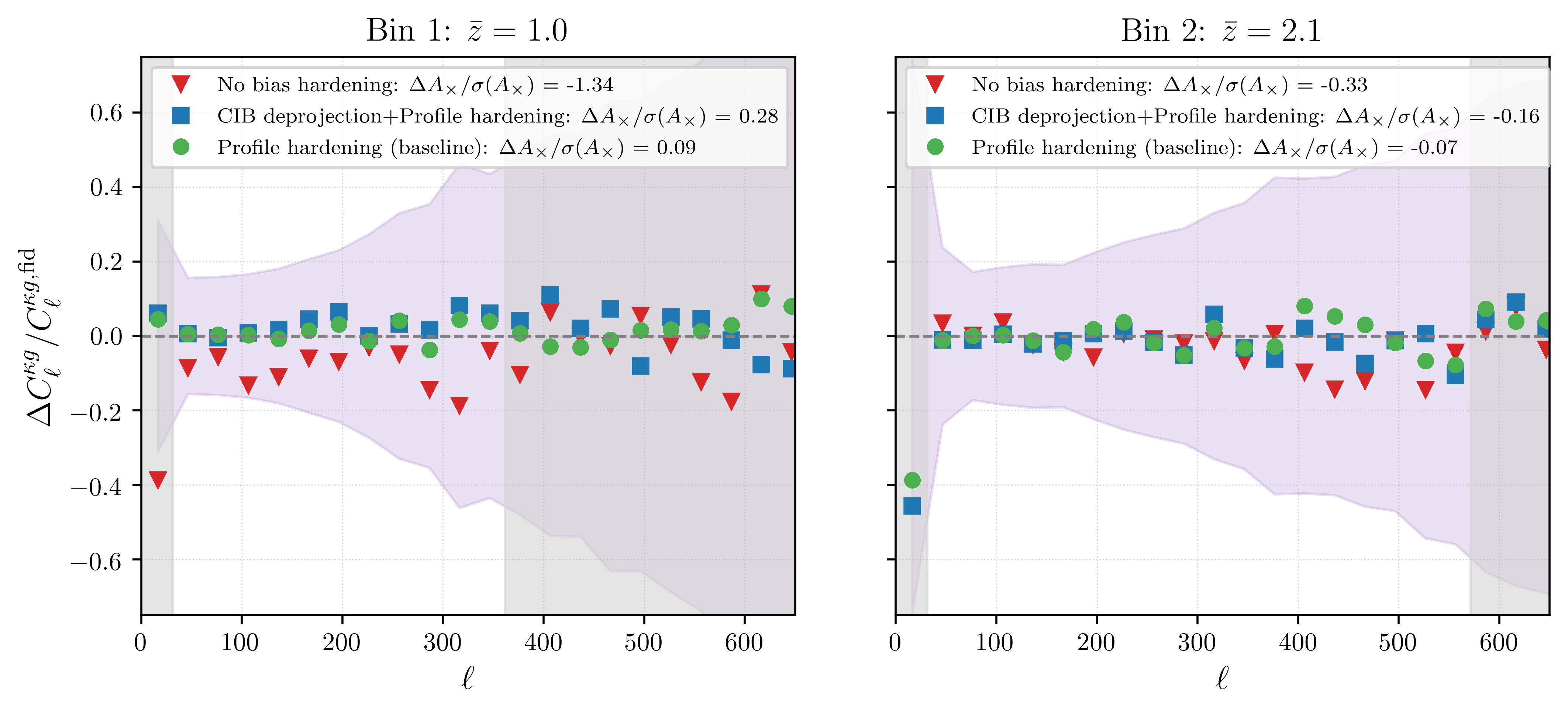}
 \caption{Cross-correlation of the lensing signal from foreground-only maps and a realization of the Quaia sample from \texttt{WebSky} populated according to the HOD
 described in Appendix \ref{app:HOD}. We compare the bias and spectra obtained with profile hardening (baseline, green circles), adding CIB deprojection (blue squares), and no geometric mitigation (red triangles), demonstrating that the baseline estimator is robust to extragalactic foreground biases. The purple bands are the $1\sigma$ errors of the measurement: $\sigma(C_\ell^{\kappa g})/C_\ell^{\kappa g\text{,fid}}$. }
 \label{fig:extragal2}
\end{figure*}

We find that residual foreground biases are not statistically significant in our baseline analysis across all redshift bins, with all values of $\Delta A_{\times}/\sigma\left(A_{\times}\right)$ well below 0.1. In Figure~\ref{fig:extragal2}, we illustrate the bias comparison evaluating three different mitigation strategies: our baseline approach using profile hardening, an alternative estimator with CIB deprojection, and a naive estimator with no bias hardening. The results highlight the substantial improvement when using the profile-hardening method \citep{Namikawa_2013,Osborne_2014,Sailer2020, Sailer2023}.

Based on these findings, we conclude that extragalactic foreground biases in the lensing maps can be safely neglected in our analysis. While previous cross-correlation  studies with the Quaia quasars have suggested potential foreground contamination in the \textit{Planck} CMB lensing maps \citep{alonso2023}, we find no evidence for such contamination in the ACT DR6 lensing maps, based on the null tests and simulation-based tests presented in this section. We note that, in any case, this potential contamination of the \textit{Planck} lensing maps was found not to impact cosmological constraints \citep{Piccirilli:2024xgo}.

\section{Cosmological analysis} \label{sec:cosmoanal}

In this section, we outline the blinding procedure adopted to mitigate the effect of confirmation bias (Section~\ref{sec:cosmoanal-blinding}), describe the likelihood and priors used in our cosmological inference (Section~\ref{sec:cosmoanal-likelihood}), and present the external baryon acoustic oscillation (BAO) likelihoods used in the analysis (Section~\ref{sec:cosmoanal-BAOlikes}).

\subsection{Blinding policy} \label{sec:cosmoanal-blinding}

To mitigate confirmation bias, we followed the blinding policy described below. No comparisons between measured spectra and theoretical predictions, nor any inference of cosmological parameters, were made until the pipeline passed the validation steps outlined below. However, we were not blind to the previously measured spectra: $C_\ell^{gg}$ or $C_\ell^{\kappa g}$ using the \textit{Planck} PR4 lensing map. The tests performed before unblinding our measurement were the following:

\begin{itemize}

    \item The pipeline successfully reproduces the results presented in \citet{alonso2023}. Specifically, we recover both the quasar auto-correlation spectrum, $C_\ell^{gg}$, and its cross-correlation with \textit{Planck} PR4 CMB lensing, $C_\ell^{\kappa_{\mathrm{Planck}}g}$, along with cosmological parameter constraints consistent with those of \citet{alonso2023} when using their same analysis choices.

    \item The measurements are free from significant Galactic or extragalactic contamination. This is established through a suite of null tests (detailed in Section~\ref{sec:tests}), which are considered to pass if their PTE values exceed 0.05. We consider this requirement to be met if failures are determined to be consistent with statistical fluctuations. We do not conduct tests on the quasar auto-correlation spectrum, as these were already done extensively in \citet{alonso2023,Piccirilli:2024xgo,Fabbian:2025fdk}.

    \item When replacing the data with a binned theory prediction in the likelihood, and using the same priors, covariance, and convergence criterion as in our baseline setup, the pipeline recovers the correct cosmological parameters to within $0.2\sigma$.

    \item Before unblinding the inferred parameters obtained using the ACT DR6 lensing maps, we freeze our prior choices for both cosmological and nuisance parameters (i.e., galaxy bias and shot noise).
    
\end{itemize}

\textit{Changes post unblinding -- }During the reanalysis with \textit{Planck} PR4 CMB lensing, and after comparing the bandpowers with those obtained using ACT DR6 lensing, we observed that the measurement was less robust to variations in scale cuts than was reported in \citet{alonso2023}, particularly when allowing the shot-noise amplitude to vary (see Appendix~\ref{app:Planck-reanalysis} for details). As a result, we adopted a more conservative choice of scale cuts in our baseline analysis, from $k_{\max} = 0.22\, h\mathrm{Mpc}^{-1}$ \citep[that we had chosen to match that of ][]{alonso2023} to $k_{\max} = 0.15\,h\mathrm{Mpc}^{-1}$, which is more consistent with the validity of a linear bias model.

\subsection{Likelihood and priors} \label{sec:cosmoanal-likelihood}

We use a Gaussian likelihood ($\mathcal{L}$) to constrain the parameters of our model ($\boldsymbol{\theta}$). For the 2$\times$2pt analysis, the likelihood is given by
\begin{equation}
-2 \ln \mathcal{L} \propto  
\left[ 
\begin{array}{c}
\Delta C^{g_1g_1}_\ell (\boldsymbol{\theta}) \\
\Delta C^{\kappa g_1}_\ell (\boldsymbol{\theta}) \\
\Delta C^{g_2g_2}_\ell (\boldsymbol{\theta}) \\
\Delta C^{\kappa g_2}_\ell (\boldsymbol{\theta})
\end{array}
\right]^{T}
\mathbb{C}^{-1}
\left[
\begin{array}{c}
\Delta C^{g_1g_1}_{\ell'} (\boldsymbol{\theta}) \\
\Delta C^{\kappa g_1}_{\ell'} (\boldsymbol{\theta}) \\
\Delta C^{g_2g_2}_{\ell'} (\boldsymbol{\theta}) \\
\Delta C^{\kappa g_2}_{\ell'} (\boldsymbol{\theta})
\end{array}
\right] \; , 
\end{equation}
where $\Delta C_\ell^{g_ig_i}$ and $\Delta C_\ell^{\kappa g_i}$ represent the residuals between the binned data and their corresponding theoretical predictions, and $\mathbb{C}$ denotes the covariance matrix presented in Section~\ref{sec:covariance}. For the 3$\times$2pt analysis we add the residuals of the CMB lensing auto-spectrum prediction $\Delta C_\ell^{\kappa \kappa}$ and incorporate the relevant correlations in the covariance matrix, as described in Section~\ref{sec:covariance}. 

To ensure a valid comparison between theory and data, we account for the pixel window function corresponding to the \texttt{HEALPix} resolution of the maps ($\texttt{nside} = 512$), and convolve the theoretical predictions with the bandpower window functions derived from the masks used in each measurement. This accounts for the scale-dependent mode-coupling introduced by the masking procedure.

The priors adopted in our analysis are summarized in Table~\ref{table:priors}. Since our data are not sensitive to the optical depth to reionization, we fix $\tau$ to the best-fit value from \textit{Planck} \citep{Planck2020}, and assume a minimal neutrino mass consistent with the normal hierarchy, $\sum m_\nu = 0.06$ eV, throughout the analysis. In the 2$\times$2pt case, the spectral index of primordial fluctuations, $n_s$, is fixed to its \textit{Planck} best-fit value, and, as the cross-correlation measurement is not sensitive to the BAO feature, we do the same with the physical baryon density, $\Omega_b h^2$. The Hubble constant $H_0$, and the cold dark matter physical density $\Omega_c h^2$ are assigned the same priors used in \citet{alonso2023}, while for the amplitude of scalar perturbations $A_s$ we follow some of the recent cross-correlation analyses \citep[e.g.,][]{Farren_2024}.

The galaxy bias parameters $b_g^i$ (as defined in Equation~\ref{eq:bias}) are assigned uniform priors in the range $[0.1, 3]$ for each redshift bin. Unlike the baseline analysis in \citet{alonso2023}, where the shot-noise amplitudes were fixed to their Poisson expectations, we allow the shot-noise amplitudes to vary. Specifically, we impose Gaussian priors centered on the Poisson estimate $\hat{N}$ (see Equation \ref{eq:sn}), with a standard deviation corresponding to 10\% of the mean. Separate bias and shot-noise parameters are used for the ACT and \textit{Planck} footprints in the joint analysis to account for potential differences across regions of the sky.

In the 3$\times$2pt case, we adopt the same priors described above for both cosmological and nuisance parameters, except for the following modifications. Following \citet{Farren_2024}: we impose a Gaussian prior on $n_s$ centered on the \textit{Planck} measurement with a width five times larger than the corresponding posterior; we adopt a BBN-informed prior on $\Omega_b h^2$; and we widen the prior on $\Omega_c h^2$.

As described in Section~\ref{sec:theory-reconstruction}, we reconstruct the evolution of the growth of structure using three amplitude parameters $A_i$, for which we use uniform priors in the range $[0, 2]$. Due to the near-complete degeneracy between these parameters and the amplitude of primordial scalar perturbations, we fix $A_s$ to its \textit{Planck} best-fit value  when we reconstruct the growth of structure in Section~\ref{sec:results-reconstruction}.

\begin{table}[]
\begin{tabular}{ccc}
\toprule
\textbf{Parameter} & \textbf{Prior ($2\times 2$pt)}         & \textbf{Prior ($3\times 2$pt)}          \\ \midrule
\multicolumn{3}{c}{\textbf{Cosmological}}                                                      \\ \midrule
$n_s$             & 0.9665                          & $\mathcal{N}(0.96, 0.02)$       \\
$\Omega_b h^2$    & 0.0224                          & $\mathcal{N}(0.02233, 0.00036)$ \\
$\tau$            & 0.0561                          & 0.0561                          \\
$\sum m_\nu$      & $0.06\, \text{eV}$                            & $0.06\, \text{eV}$                            \\
$\ln(10^{10}A_s)$ & [1.0,4.0]                       & [1.0,4.0]                       \\
$H_0$             & [50,80]                         & [50,80]                         \\
$\Omega_c h^2$    & [0.08,0.20]                     & [0.005,0.99]                    \\ \midrule
\multicolumn{3}{c}{\textbf{Nuisance}}                                                          \\ \midrule
$b_1$             & [0.1,3]                           & [0.1,3]                           \\
$b_2$             & [0.1,3]                           & [0.1,3]                           \\
$A_{N1}$          & $\mathcal{N}(\hat{N}_1, \, 0.1\, \hat{N}_1)$ & $\mathcal{N}(\hat{N}_1,\, 0.1\,\hat{N}_1)$ \\
$A_{N2}$          & $\mathcal{N}(\hat{N}_2,\, 0.1\,\hat{N}_2)$ & $\mathcal{N}(\hat{N}_2,\, 0.1\,\hat{N}_2)$ \\ \midrule
\multicolumn{3}{c}{\textbf{Growth reconstruction} (Equation~\ref{eq:growth-rec})}                                      \\ \hline
$A_{1}$           & -                               & [0,2]                           \\
$A_{2}$           & -                               & [0,2]                           \\
$A_{3}$           & -                               & [0,2]   \\                  
\bottomrule
\end{tabular}
\caption{Parameters and priors used in both the 2$\times$2pt and 3$\times$2pt inference. A normal distribution is denoted by $\mathcal{N}(\mu, \sigma)$, indicating a mean $\mu$ and standard deviation $\sigma$. Uniform priors are specified by ranges in square brackets. Parameters listed with a single value are held fixed during the analysis. The mean values for the shot-noise prior $\hat{N}_1$ and $\hat{N}_2$ can be found on Table \ref{table:Quaia}. Note that when we open up the parameter space with $A_i$ to reconstruct the growth of structure across redshifts, we fix $A_s$ to its \textit{Planck} best-fit value ($\ln(10^{10} A_s) = 3.041$;~\citealt{Planck2020}).}
\label{table:priors}
\end{table}

Parameter inference is performed using the Markov Chain Monte Carlo (MCMC) framework implemented in \texttt{Cobaya} \citep{Torrado_2021}. Chains are considered to be converged when the Gelman--Rubin statistic \citep{Rubin, Brooks} satisfies $R-1 < 0.01$.

\subsection{Baryon acoustic oscillation likelihoods} \label{sec:cosmoanal-BAOlikes}

We incorporate BAO measurements to help break the degeneracies between $\Omega_m$, $\sigma_8$, and $H_0$. The bulk of the analysis presented in this work was completed prior to the public release of the DESI DR2 BAO data. Therefore, the majority of measurements and tests in this work rely on BAO data from the 6dF and SDSS surveys, which we collectively refer to as simply ``BAO'' throughout. Specifically, we include the following likelihoods: 6dFGS \citep{6dFGS}, the SDSS DR7 Main Galaxy Sample \citep{MGS}, BOSS DR12 Luminous Red Galaxies (LRGs) \citep{LRG12}, and eBOSS DR16 LRGs \citep{LRG16}, Emission Line Galaxies (ELGs) \citep{ELG}, Lyman-$\alpha$ forest \citep{lyman}, and quasars \citep{quasarBAO}.

We also revisit several of our key results using the recently released DESI DR2 BAO measurements \citep{desicollaboration2025desidr2resultsii}, given their significantly improved sensitivity to cosmological parameters (which we refer to as ``DESI BAO'').

\section{Results} \label{sec:results}

We organize the results in three main parts. Section \ref{sec:results-2x2pt} summarizes the constraints obtained from the 2$\times$2pt analysis, i.e., using the quasar auto‑spectra $C_\ell^{gg}$ and their cross‑correlation with CMB lensing, $C_\ell^{\kappa g}$. In Section \ref{sec:results-3x2pt} we add the CMB‑lensing auto‑spectrum $C_\ell^{\kappa\kappa}$ to update cosmological parameters using the 3$\times$2pt combination before performing a reconstruction of structure growth across redshifts in Section \ref{sec:results-reconstruction}. We report the results in this section quoting the mean of the posterior and 1$\sigma$ standard deviation.

\subsection{2$\times$2pt cosmology} \label{sec:results-2x2pt}

Here, we present the cosmological constraints from the cross-correlation of Quaia quasars and ACT DR6 CMB lensing (Section~\ref{sec:results-2x2pt-ACT}), along with the joint constraints when adding \textit{Planck} PR4 (Section~\ref{sec:results-2x2pt-Planck}). A summary of all cosmological constraints presented in this section is provided in Table~\ref{table:results}.

\begin{table}[h!]
    \centering
    \renewcommand{\arraystretch}{1.3}
    \setlength{\tabcolsep}{8pt}
    \begin{tabular}{cccc}
        \toprule
        & $\Omega_m$ & $\sigma_8$ & $S_8$ \\
        \midrule
        \multicolumn{4}{c}{\textbf{ACT DR6 $\times$ Quaia}} \\
        \midrule
        Bin 1 & $0.405^{+0.075}_{-0.14}$ & $0.84^{+0.13}_{-0.21}$ & $0.95^{+0.10}_{-0.22}$ \\
        Bin 2 & $0.397^{+0.080}_{-0.14}$ & $0.631^{+0.069}_{-0.11}$ & $0.713^{+0.084}_{-0.13}$ \\
        Joint & $0.411^{+0.084}_{-0.12}$ & $0.685^{+0.074}_{-0.11}$ & $0.787^{+0.067}_{-0.099}$ \\
        \midrule
        \multicolumn{4}{c}{\textbf{ACT DR6 $\times$ Quaia + BAO}} \\
        \midrule
        Bin 1 & $0.313^{+0.017}_{-0.019}$ & $0.90^{+0.12}_{-0.18}$ & $0.92^{+0.12}_{-0.18}$ \\
        Bin 2 & $0.313 \pm 0.018$ & $0.650^{+0.065}_{-0.11}$ & $0.664^{+0.066}_{-0.11}$ \\
        Joint & $0.313^{+0.016}_{0.019}$ & $0.744^{+0.064}_{-0.92}$ & $0.759^{+0.065}_{-0.094}$ \\
        \midrule
        \multicolumn{4}{c}{\textbf{(ACT DR6 + \textit{Planck} PR4) $\times$ Quaia}} \\
        \midrule
        Bin 1 & $0.477^{+0.094}_{-0.14}$ & $0.723^{+0.085}_{-0.13}$ & $0.895^{+0.073}_{-0.12}$ \\
        Bin 2 & $0.367^{+0.066}_{-0.12}$ & $0.761^{+0.073}_{-0.096}$ & $0.829^{+0.083}_{-0.11}$ \\
        Joint & $0.432^{+0.095}_{-0.11}$ & $0.733^{+0.061}_{-0.094}$ & $0.865^{+0.062}_{-0.078}$ \\
        \midrule
        \multicolumn{4}{c}{\textbf{(ACT DR6 + \textit{Planck} PR4) $\times$ Quaia + BAO}} \\
        \midrule
        Bin 1 & $0.327^{+0.017}_{-0.019}$ & $0.825^{+0.067}_{-0.10}$ & $0.861^{+0.069}_{-0.11}$ \\
        Bin 2 & $0.312^{+0.016}_{-0.017}$ & $0.783^{+0.062}_{-0.084}$ & $0.797^{+0.064}_{-0.088}$ \\
        Joint & $0.322^{+0.014}_{-0.016}$ & $0.802^{+0.045}_{-0.057}$ & $0.830^{+0.047}_{-0.059}$ \\
        \bottomrule
    \end{tabular}
    \caption{
    Summary of cosmological parameter constraints on $\Omega_m$, and $\sigma_8$ and $S_8$, at redshift $z=0$ derived from cross-correlations (2$\times$2pt) between Quaia quasars and both ACT DR6 and \textit{Planck} PR4 CMB lensing maps. We present results separately for each Quaia sample, their combination, and also include constraints obtained with external BAO measurements from 6dF and BOSS.}
    \label{table:results}
\end{table}

\subsubsection{ACT DR6 $\times$ Quaia} \label{sec:results-2x2pt-ACT}

\begin{figure}
 \centering
 \includegraphics[width=\linewidth]{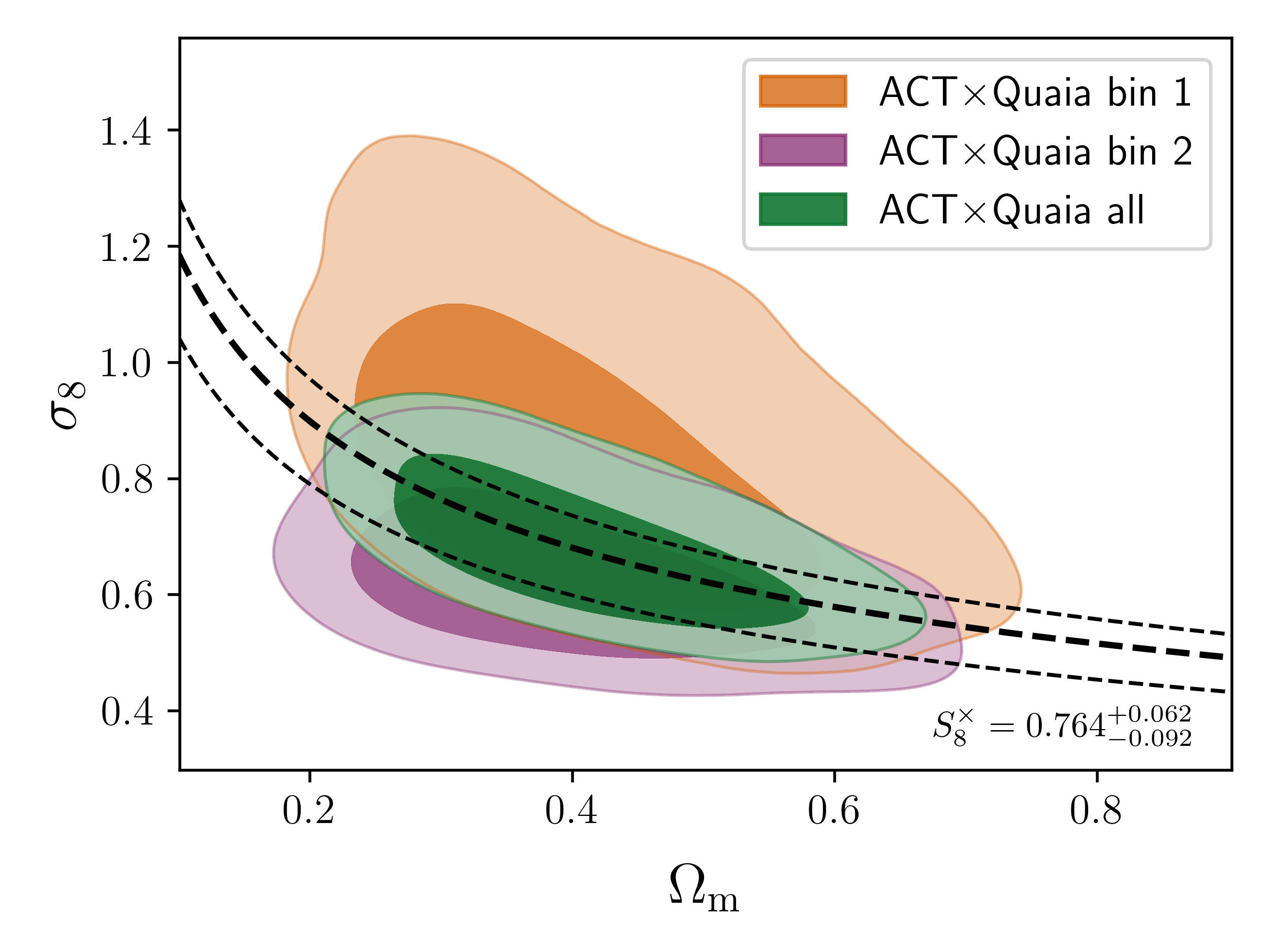}
 \caption{Constraints obtained from the cross-correlation of ACT DR6 CMB lensing with the Quaia quasar samples, shown as 1$\sigma$ and 2$\sigma$ confidence contours in the $\sigma_8$–$\Omega_m$ plane. The dashed black lines represent the best-fit value of the parameter $S_8^{\times}$ and its associated 1$\sigma$ uncertainty.}
 \label{fig:ACTxQuaia}
\end{figure}

We first consider the combination of the quasar auto-correlation measured within the ACT footprint and its cross-correlation with ACT DR6 CMB lensing. Since these observables constrain a combination of $\sigma_8$ and $\Omega_m$, we report constraints in terms of $S_8 \equiv \sigma_8 (\Omega_m/0.3)^{0.5}$. We obtain a roughly $10\%$ constraint giving
\begin{equation*}
    S_8 = 0.787^{+0.067}_{-0.099}  \; \text{(ACT DR6 $\times$ Quaia)} \; .
\end{equation*}
Our data most strongly constrain a slightly different parameter combination, which we define as $S_8^{\times} \equiv  \sigma_8 (\Omega_m/0.3)^{0.4}$ finding a value of $S_8^{\times} = 0.764^{+0.062}_{-0.092}$. Figure~\ref{fig:ACTxQuaia} shows the joint posterior distribution in the $\sigma_8$–$\Omega_m$ plane for the different Quaia samples as well as the mean value and $1\sigma$ uncertainty for our best-constrained parameter $S_8^{\times}$.

We break the degeneracy between $\sigma_8$ and $\Omega_m$ by adding external BAO information. Using the combination of BAO likelihoods described in Section~\ref{sec:cosmoanal-BAOlikes} -- which primarily constrains $\Omega_m$ -- allows us to measure $\sigma_8$ independently. Combining the two Quaia redshift bins, we achieve an approximately 10\% constraint on $\sigma_8$:
\begin{equation*}
    \sigma_8 = 0.744^{+0.064}_{-0.092}  \; \text{(ACT DR6$\times$Quaia + BAO)} \; .
\end{equation*}
Repeating this analysis with the more recent DESI DR2 BAO data described in Section~\ref{sec:cosmoanal-BAOlikes}, we find a consistent constraint of $\sigma_8 = 0.762^{+0.061}_{-0.10}$.

\begin{figure*}
 \centering
 \includegraphics[width=\linewidth]{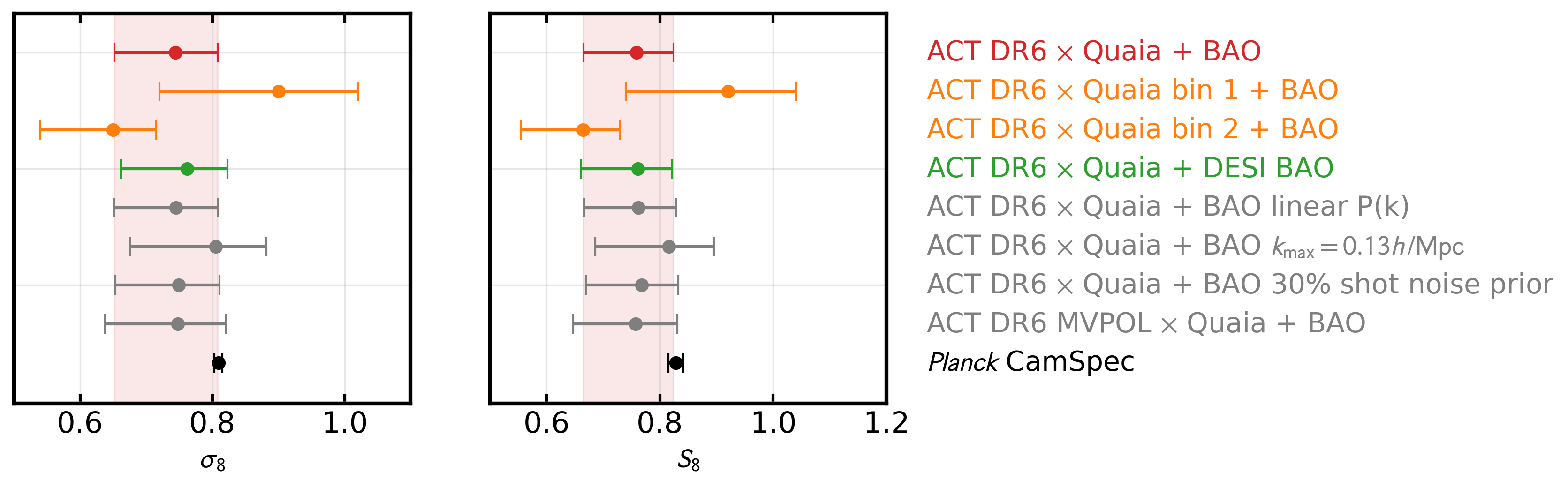}
 \caption{Summary of constraints on $\sigma_8$ (left) and $S_8$ (right) obtained from different data combinations and analysis choices. The baseline result from ACT DR6 and both Quaia redshift samples combined with BOSS and 6dF BAO, is highlighted in red and with a shaded band. Results from analyzing each Quaia redshift sample separately, are shown in orange. Constraints derived using DESI DR2 BAO measurements are displayed in green. Grey points represent various consistency tests performed to assess robustness to analysis choices. For comparison, constraints from the primary CMB anisotropy analysis  \citep[\textit{Planck} PR3 CamSpec likelihood;][]{Planck2020} are shown in black.}
 \label{fig:ACTxQuaia_tests}
\end{figure*}

We demonstrate that these results are robust with respect to various analysis choices by performing a suite of consistency tests, summarized in Figure~\ref{fig:ACTxQuaia_tests}. Each test, shown in grey, is consistent within $1\sigma$ of our baseline measurement. In particular, we highlight the agreement when performing the analysis on the polarization-only convergence maps (MVPOL $\kappa$ described in Section~\ref{sec:null-tests}), which, as mentioned above, is expected to be insensitive to extragalactic foreground contamination.

Previous analyses \citep{alonso2023,Piccirilli:2024xgo} pointed to ``weak evidence'' towards extragalactic contamination (such as CIB) in the \textit{Planck} CMB lensing maps based on mainly two observations: 
\begin{itemize}
    \item The inferred value of $\sigma_8$ from cross-correlating \textit{Planck} lensing with the Quaia high-redshift sample was found to be lower than the prediction from \textit{Planck} primary CMB by 2.5$\sigma$.
    \item The cross-correlation measurement using polarization-only lensing maps showed a relatively coherent upward shift compared to the minimum-variance (MV) reconstruction, yielding a slightly higher value of $\sigma_8$ (approximately $1.1\sigma$ above the baseline result obtained with the MV map). While the statistical uncertainty was larger, this measurement was in better agreement with \textit{Planck} primary CMB predictions.
\end{itemize}
In contrast, when performing our analysis using ACT lensing, we find consistent values of $\sigma_8$ when comparing our baseline minimum-variance maps to polarization-only reconstructions. Furthermore, our extensive set of null and foreground-contamination tests, presented in Section~\ref{sec:tests}, show no evidence of CIB or other extragalactic contamination in the ACT DR6 lensing data.

Nonetheless, we still find a slightly low value of $\sigma_8$ (at around the $1.8\sigma$ level) in the higher redshift bin compared to primary CMB. If this deviation shares the same origin with the earlier \textit{Planck} cross-correlation, it may point to a feature intrinsic to the Quaia sample -- common to both analyses. However, this may not necessarily be the case: these shifts could arise from unrelated effects in each case, or simply reflect a statistical fluctuation given its modest significance.

\subsubsection{Combination of ACT DR6 and \textit{Planck} PR4 CMB lensing $\times$ Quaia} \label{sec:results-2x2pt-Planck}

We present an updated analysis of the cross-correlation between the Quaia quasar sample and \textit{Planck} PR4 lensing maps in Appendix~\ref{app:Planck-reanalysis}, where we demonstrate the necessity of adopting more conservative scale cuts, as well as the importance of allowing the shot-noise amplitudes to vary. Using these updated analysis choices in combination with BAO we obtain a $7\%$ constraint $\sigma_8 = 0.797^{+0.048}_{-0.063}$, consistent with the measurement using ACT lensing at the $0.45\sigma$ level.

Despite the lower reconstruction noise in ACT, we find that the cross-correlation measurement using \textit{Planck} lensing has a higher SNR than that of ACT, primarily due to its larger overlap with the Quaia footprint. Moreover, the auto-correlation of the quasar overdensity measured over the full \textit{Planck} footprint has a higher SNR than when measured only on the ACT footprint. This explains why the constraints on $S_8$ and $\sigma_8$ from \textit{Planck} PR4 are tighter than those using ACT DR6 lensing alone.

Given the consistency between constraints derived from ACT DR6 and \textit{Planck} PR4 lensing data, we perform a joint analysis combining their cross-correlations information. As detailed in Section~\ref{sec:cosmoanal-likelihood}, we use different bias and shot-noise parameters for the ACT and \textit{Planck} footprints. The covariance matrix for this joint analysis includes correlations between $C_\ell^{\kappa_{\text{ACT}}g}$ and $C_\ell^{\kappa_{\text{Planck}}g}$, and is described in Section~\ref{sec:covariance}. We additionally incorporate the correlations between the quasar auto-correlations on different footprints ($C_\ell^{gg,\text{ACT}}$ and $C_\ell^{gg,\text{Planck}}$) from the Gaussian simulations described in Section~\ref{sec:simulations}.

The joint analysis yields an 8\% constraint on $S_8 = 0.865^{+0.062}_{-0.078}$.
We also report this result in terms of our best-constrained parameter for which we find a 7\% constraint: $S_8^{\times} = 0.836^{+0.053}_{-0.070}$.

\begin{figure*}
  \centering
  \begin{minipage}[t]{0.48\textwidth}
    \centering
    \includegraphics[width=\linewidth]{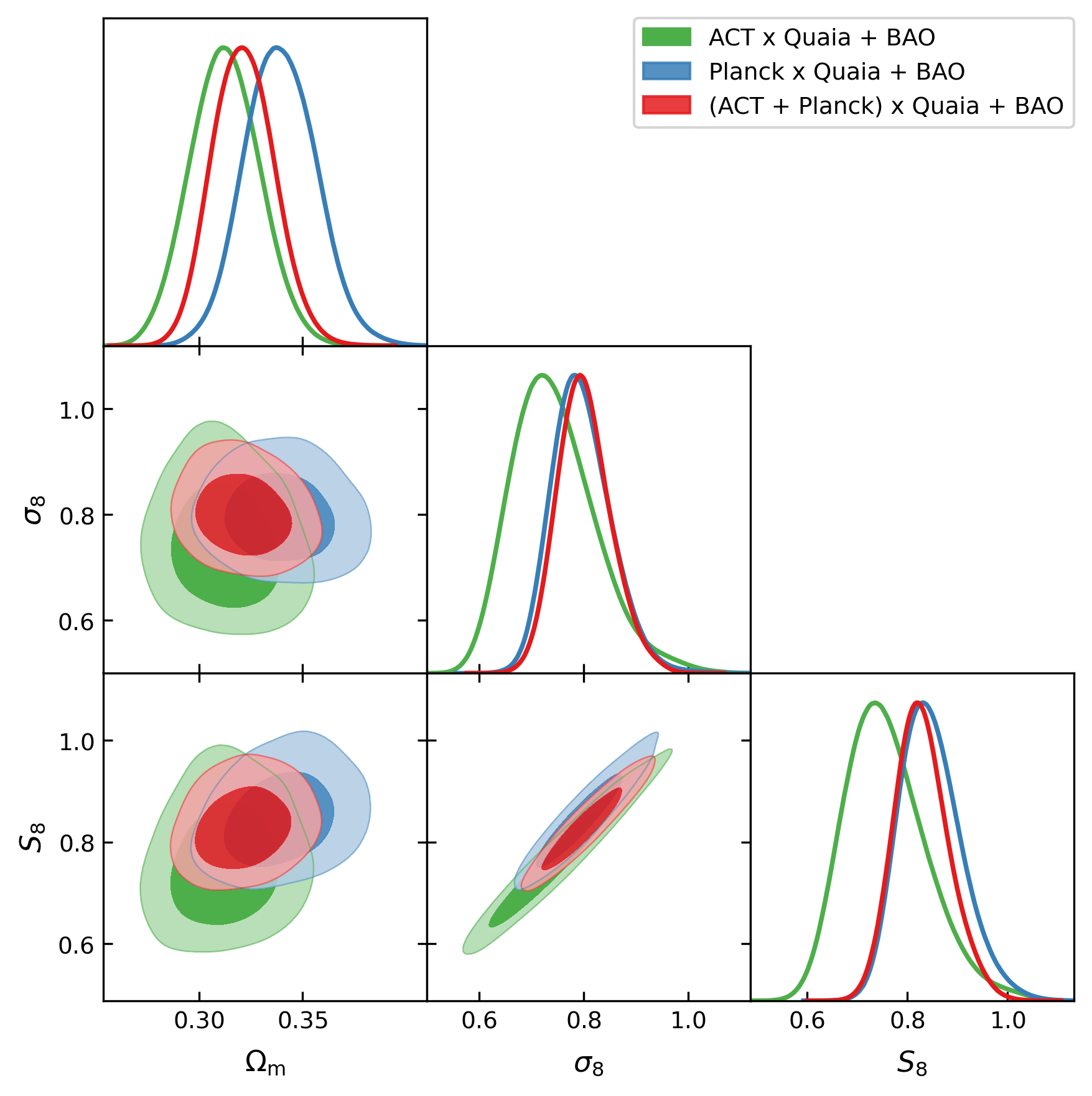}
  \end{minipage}\hfill
  \begin{minipage}[t]{0.48\textwidth}
    \centering
    \includegraphics[width=\linewidth]{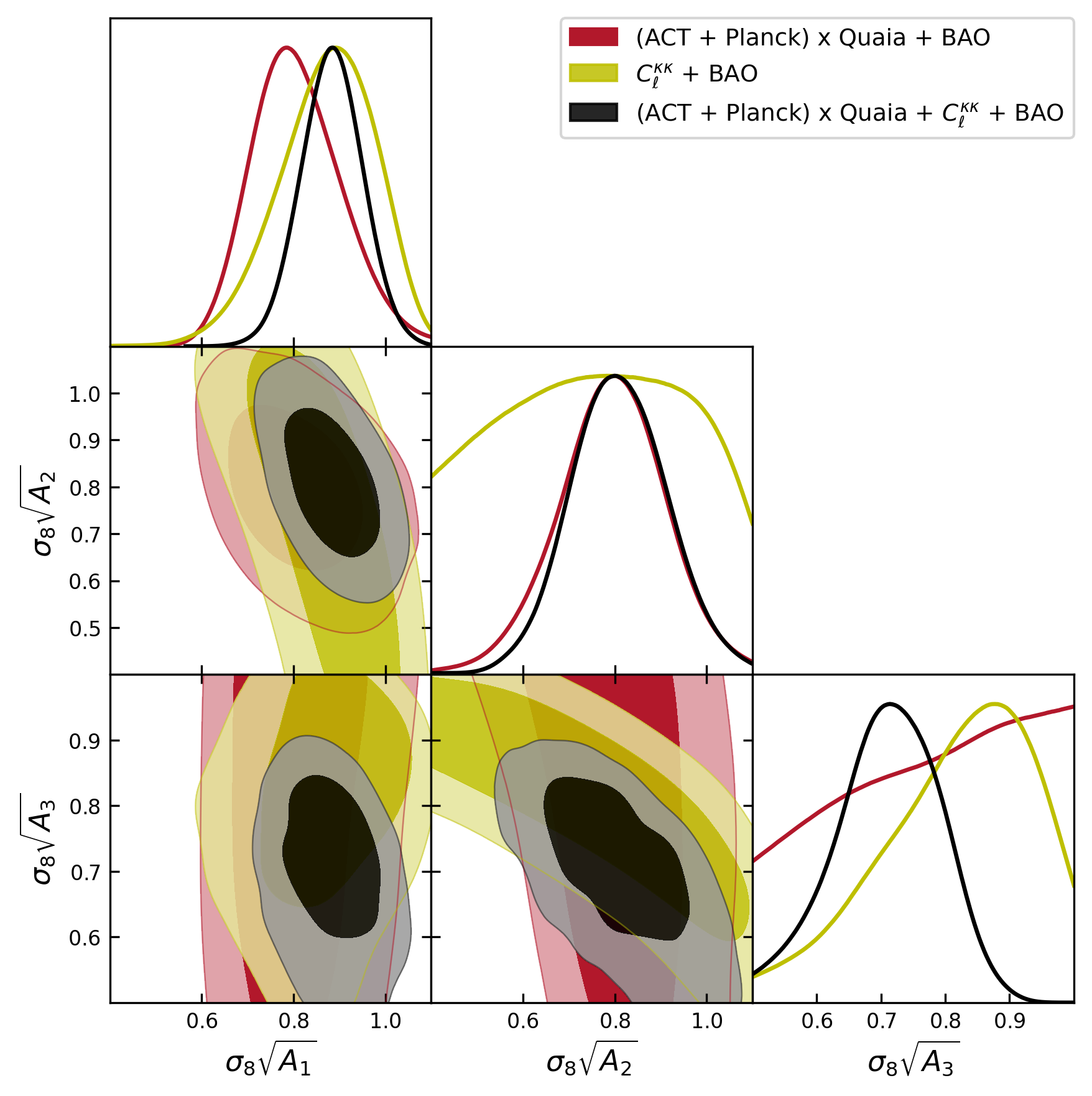}
  \end{minipage}
  \caption{\textit{Left}: Joint constraints obtained from the cross-correlation of ACT DR6 lensing (green), \textit{Planck} PR4 lensing (blue), and their combination (red) with the Quaia quasar samples, including BAO information described in Section \ref{sec:cosmoanal-BAOlikes}. We show the 1$\sigma$ and 2$\sigma$ confidence contours. \textit{Right}: Constraints on the relevant amplitude of structure growth parameters $\sigma_8 \sqrt{A_1}$, $\sigma_8 \sqrt{A_2}$, and $\sigma_8 \sqrt{A_3}$ from the 2$\times$2pt (red), $C_\ell^{\kappa \kappa}$-only (yellow) and 3$\times$2pt (black) analyses. The amplitudes $A_i$ rescale the matter power spectrum in three redshift intervals: $z<1.45$, $1.45\leq z<3.0$, and $3.0\leq z$, as described in Section \ref{sec:theory-reconstruction}, enabling us to constrain the evolution of structure growth beyond the redshift range directly probed by the quasars.}
  \label{fig:ACT+PlanckxQuaia+BAO}
  \label{fig:kk_separate_results}
\end{figure*}

When incorporating BAO information, the joint analysis gives approximately a 6.4\% constraint:
\begin{equation*}
\sigma_8 = 0.802^{+0.045}_{-0.057}\;\left[\text{(ACT+\textit{Planck})$\times$Quaia + BAO}\right] \; .
\end{equation*}
The comparison between the ACT and \textit{Planck} cross-correlation measurements, as well as their combination, is shown in Figure \ref{fig:ACT+PlanckxQuaia+BAO}.

We note that the combined constraint on $\sigma_8$ is slightly higher than those from ACT and \textit{Planck} individually --by approximately 0.1$\sigma$ compared to the higher of the two (from \textit{Planck}). This increase arises because the ACT cross-correlation prefers slightly lower values of $\Omega_m$, shifting the joint $\Omega_m$–$\sigma_8$ contour in that direction and resulting in a marginally higher $\sigma_8$ when combined with BAO. We also repeat this analysis using DESI BAO, which gives a slightly (less than $0.25\sigma$) higher $\sigma_8$ value: 
\begin{equation*}
\sigma_8 =0.814^{+0.045}_{-0.056}\;\left[\text{(ACT+\textit{Planck})$\times$Quaia + DESI BAO}\right] \; .
\end{equation*}

\subsection{3$\times$2pt cosmology} \label{sec:results-3x2pt}

We now report the cosmological constraints obtained when including the CMB lensing power spectrum $C_{\ell}^{\kappa \kappa}$ in our analysis, extending the 2$\times$2pt framework to the full 3$\times$2pt case. The corresponding updates to the covariance matrix and prior choices are described in Sections \ref{sec:covariance} and \ref{sec:cosmoanal-likelihood}, respectively.

We obtain a constraint on structure growth of $S_8 = 0.875^{+0.044}_{-0.029}$. For this data combination, the best-constrained parameter is given by $S_8^{0.3} \equiv \sigma_8 (\Omega_m/0.3)^{0.3}$ for which we find a 1.6\% constraint $S_8^{0.3} = 0.803 \pm 0.013$.

We further combine this data with the BAO measurements described in Section~\ref{sec:cosmoanal-BAOlikes}, leading to a 1.6\% constraint on $\sigma_8$:
\begin{equation*}
    \sigma_8 = 0.804 \pm 0.013 \; \left[\text{(ACT+Planck)$\times$Quaia+$C_\ell^{\kappa \kappa}$+BAO}\right]\; ,
\end{equation*}
finding it to be fully consistent with \textit{Planck} primary CMB predictions. While the dominant contribution for this constraint is coming from $C_\ell^{\kappa \kappa}$, we find that the inclusion of the 2$\times$2pt information leads to a variance cancellation, tightening the constraint on $\sigma_8$ by approximately 12\% relative to $C_\ell^{\kappa \kappa}$ alone. Repeating the analysis with DESI BAO yields $\sigma_8 = 0.810 \pm 0.012$. As expected, the $\sigma_8$ values inferred using DESI BAO are slightly higher than those obtained with BOSS and 6dF BAO, reflecting DESI’s preference for a lower value of $\Omega_m$.

\subsection{Growth-of-structure reconstruction} \label{sec:results-reconstruction}

As described in Section~\ref{sec:theory-reconstruction}, we exploit the different redshift sensitivity of our measurements to constrain the evolution of structure growth across redshifts. While the redshift kernel of the lensing power spectrum $C_\ell^{\kappa\kappa}$ has its median at $z \sim 2$, it receives significant contributions from lower redshifts that overlap with the Quaia samples. This overlap enables us to extract additional information on the amplitude of structure growth at redshifts beyond those directly probed by the quasars.

To extract this information, we adopt the parametrization presented in Section~\ref{sec:theory-reconstruction}, rescaling the matter power spectrum with three amplitude parameters $A_1$, $A_2$, and $A_3$. These correspond to the redshift intervals $0\lesssim z\lesssim z_1$, $z_1\lesssim z\lesssim z_2$, and $z_2\lesssim z$, respectively, with $z_1 = 1.45$ and $z_2 = 3.0$ chosen approximately to match the redshift kernels of the Quaia samples and minimize correlations between redshift bins. This parametrization isolates the contributions from the redshift ranges probed by the 2$\times$2pt, allowing the lensing auto-spectrum $C_\ell^{\kappa\kappa}$ to constrain structure growth at $z \gtrsim 3.0$. 

We determine the median redshift of the signal within each redshift interval by computing the derivative of the signal-to-noise ratio with respect to redshift $d\text{SNR}/dz$, for the 3$\times$2pt measurement. The resulting $\sigma_8$ constraints evaluated at the median redshifts $\tilde{z}$ of the $d\text{SNR}/dz$ distribution within each of the intervals are:
\begin{align*}
    \sigma_8(\tilde{z}=1.0) \sqrt{A_1} = 0.638^{+0.041}_{-0.040} \; ,\\
    \sigma_8(\tilde{z}=2.1) \sqrt{A_2} = 0.327^{+0.040}_{-0.049} \; , \\
    \sigma_8(\tilde{z}=5.1) \sqrt{A_3} = 0.146^{+0.021}_{-0.014} \; .
\end{align*}
Figure~\ref{fig:kk_separate_results} shows the constraints on these amplitude parameters using the different data combinations: 2$\times$2pt, $C_\ell^{\kappa\kappa}$-only, and 3$\times$2pt, highlighting the power of the 3$\times$2pt combination in constraining the highest redshift amplitude. We note there are residual correlations between the amplitude parameters, with correlation coefficients of 62\% between bins 2 and 3, 48\% between bins 1 and 2, and 24\% between bins 1 and 3. 

\begin{figure*}
 \centering
 \includegraphics[width=\linewidth]{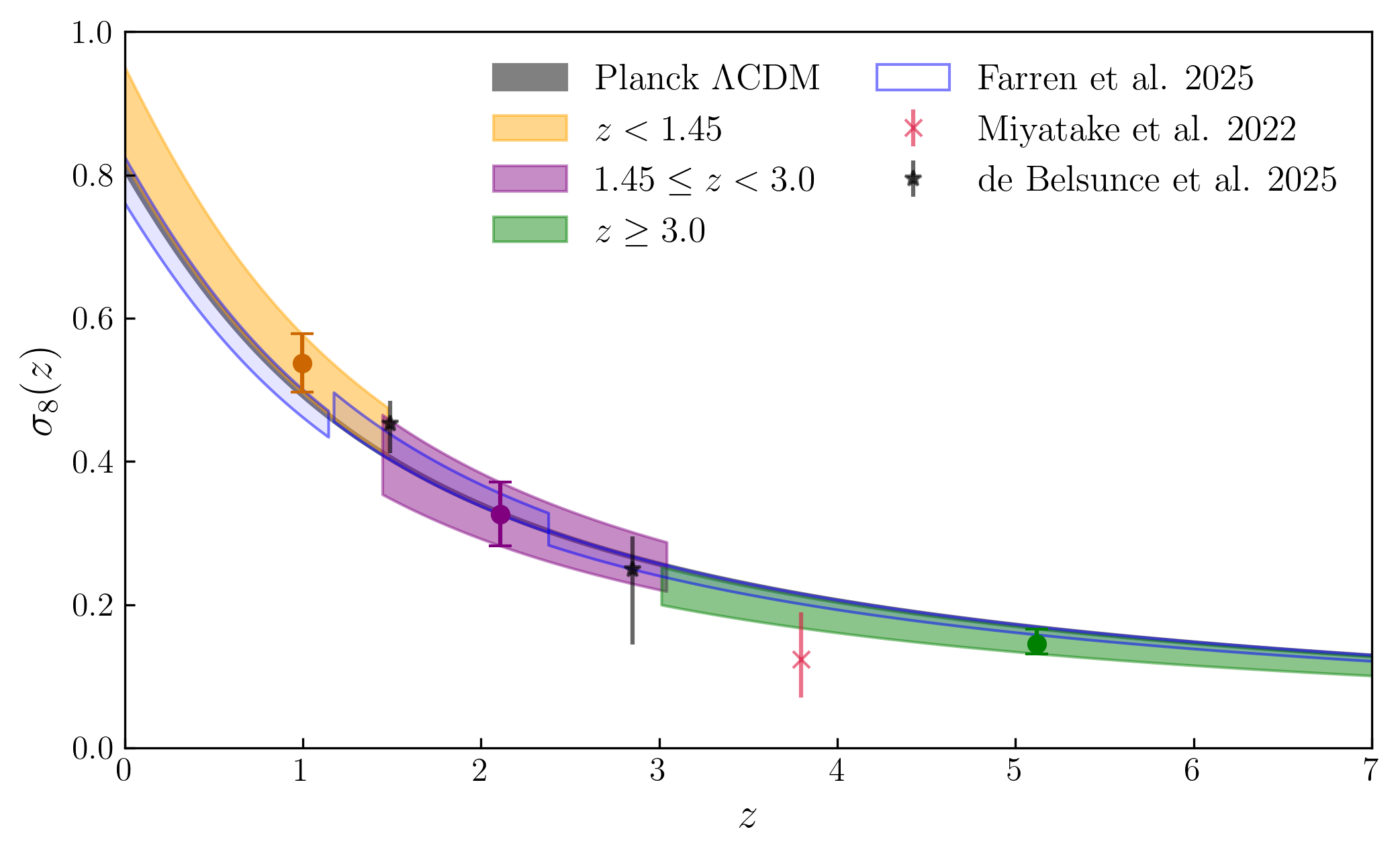}
 \caption{Reconstruction of $\sigma_8(z)$ from the 3$\times$2pt analysis. The fiducial $\sigma_8(z)$ curve is rescaled in each redshift bin by $\sqrt{A_i}$, where $A_i$ are the amplitude parameters introduced in Equation~\eqref{eq:growth-rec}. Shaded regions denote $1\sigma$ confidence intervals. This reconstruction probes the redshift evolution of the amplitude of the growth of structure out to $z > 3$. Constraints from a similar 3$\times$2pt analysis \citep{Farren_2024rla} are shown as blue bands, while the measurements from \citet{Miyatake:2021qjr} and \citet{debelsunce2025cosmologyplanckcmblensing} are shown with red and black data points, respectively. We also show the prediction from $\Lambda$CDM as measured by \textit{Planck} primary CMB \citep{Planck2020} plotted as a gray band.}

 \label{fig:sigma8(z)}
\end{figure*}

We also present a direct reconstruction of $\sigma_8(z)$ by rescaling the fiducial $\sigma_8(z)$ curve with $\sqrt{A_i}$ in each redshift interval. The result is shown in Figure~\ref{fig:sigma8(z)}, with shaded bands denoting the $1\sigma$ uncertainties, and data points highlighting the measurement at the median redshift of the signal in each interval. The overall shape of the $\sigma_8(z)$ curve follows the cosmology presented in Section \ref{sec:results-3x2pt}. Figure~\ref{fig:sigma8(z)} also compares our constraints on $\sigma_8(z)$ to the $\Lambda$CDM prediction from \textit{Planck} primary CMB, with our highest-redshift measurement consistent at the 1.4$\sigma$ level. We show results from other high-redshift measurements of structure growth by \citet{Farren_2024rla}, \citet{debelsunce2025cosmologyplanckcmblensing}, and \citet{Miyatake:2021qjr}. 

Our results are in excellent agreement with the 3$\times$2pt measurement from \citet{Farren_2024rla}, which combines \textit{unWISE} galaxies with ACT DR6 and \textit{Planck} PR4 CMB lensing. Their analysis is methodologically similar to ours, utilizing the same CMB lensing maps in the cross-correlation, and additionally incorporating the \textit{Planck} PR4 auto-spectrum $C_\ell^{\kappa \kappa}$. The main differences lie in the redshift coverage of the \textit{unWISE} galaxies and their use of smaller scales (down to $k \sim 0.3\, h\mathrm{Mpc}^{-1}$), modeled using a second-order hybrid-perturbation-theory approach for the galaxy auto- and cross-power spectra.

\citet{Miyatake:2021qjr} constrains cosmological parameters by stacking the \textit{Planck} PR3 CMB lensing maps on Lyman-break galaxies observed by the HSC Strategic Survey Program. Their analysis includes angular scales of $6'<\theta<20'$ for the lensing signal and models the convergence profile using an HOD prescription. To add their result to Figure \ref{fig:sigma8(z)}, we extrapolate their reported $\sigma_8(z=0)$ value to the mean redshift of their measurement ($z = 3.8$) using the cosmology derived in Section~\ref{sec:results-3x2pt}, as their analysis does not constrain $\Omega_m$ and therefore cannot independently determine the shape of $\sigma_8(z)$. Their result is consistent with ours within $1.1\sigma$. 

Lastly, we include the measurement from \citet{debelsunce2025cosmologyplanckcmblensing}, which cross-correlates DESI DR1 quasars with \textit{Planck} PR4 CMB lensing. The DESI quasars have a similar redshift coverage and number density to those in Quaia. However, their analysis uses smaller scales, with roughly $k_{\text{max}} = 0.21\,h\text{Mpc}^{-1}$, and is modeled with hybrid perturbation theory. Their constraints are consistent with ours within 1$\sigma$; we show their results for the first and last redshift bins, omitting the middle bin (as done in their paper) due to its limited constraining power on its own.

\section{Conclusion} \label{sec:conclussion}

We have presented new constraints on structure growth from the cross-correlation of quasars in the Quaia catalog and state-of-the-art CMB lensing maps from ACT DR6 and \textit{Planck} PR4. We also performed a 3$\times$2pt analysis by incorporating the ACT DR6 CMB lensing auto-spectrum into the cross-correlation measurement, allowing us to place one of the highest redshift constraints on structure growth to date.

Using the cross-correlation information alone of the Quaia quasars and combined ACT and \textit{Planck} lensing, we obtain a 7\% constraint on our best-constrained parameter $S^{\times}_8 = \sigma_8 (\Omega_m / 0.3)^{0.4} = 0.836^{+0.053}_{-0.070}$, and a 6.4\% constraint on $\sigma_8 = 0.802^{+0.045}_{-0.057}$ when BAO information from BOSS and 6dF is included (see Section~\ref{sec:results-2x2pt-Planck}).

We revisit the previous measurement \citep{alonso2023}, which cross-correlated the Quaia quasars with \textit{Planck} PR4 lensing only. We restrict our analysis to more conservative scale cuts and add a Gaussian prior on the shot-noise amplitude. These decisions (discussed in Appendix \ref{app:Planck-reanalysis}) are motivated by observed instabilities in the inferred cosmological parameters when using the baseline analysis choices from the previous measurement. These changes come at the cost of a SNR reduction, and, therefore, the resulting cosmological constraints are less competitive than we initially hoped. Nonetheless, this work represents an improvement on the robustness of the measurement.

To validate the robustness of our measurements with ACT lensing, we perform a suite of systematic and null tests. These include bandpower null tests using different lensing reconstructions and a foreground contamination test using simulated maps based on a HOD prescription for Quaia, described in Appendix \ref{app:HOD}. These tests show no significant evidence of contamination in the lensing maps. We also find no significant shift in our measured $\sigma_8$ parameter when performing the analysis using a polarization-only lensing reconstruction, contrary to what was found in the previous cross-correlation with \textit{Planck} \citep{alonso2023}.

Nevertheless, we recover a slightly low value of $\sigma_8$ in the high-redshift bin (discrepant with primary CMB at the 1.8$\sigma$ level) that was also found in the previous analysis using \textit{Planck} lensing. This, and a higher amplitude on the cross-correlation measurement using a polarization-only lensing reconstruction, were the motivations for suggesting ``mild evidence'' towards foreground contamination in the \textit{Planck} PR4 lensing maps \citep{Carron_2022}. However, we argue in Section \ref{sec:results-2x2pt-ACT} that, if the same in origin, this discrepancy cannot arise from contamination in the lensing maps. In that case, this deviation could reflect a feature intrinsic to the quasar sample -- which is common to both measurements -- or simply a statistical fluctuation given its low significance. 

We then extend the analysis by including the CMB lensing power spectrum $C_\ell^{\kappa\kappa}$ to perform a 3$\times$2pt measurement. This leads to a 1.6\% constraint on $\sigma_8 = 0.804\pm0.013$ when combined with BAO data, completely consistent with predictions from the primary CMB. While most of the constraining power comes from $C_\ell^{\kappa\kappa}$ alone, we also find an improvement due to variance cancellation when adding $C_\ell^{\kappa\kappa}$ to the cross-correlation analysis.

With this combined dataset, we also reconstruct the redshift evolution of structure growth by introducing a parametrization of the matter power spectrum with amplitude parameters defined in different redshift intervals. The cross-correlation with quasars constrains the low-redshift contribution to the lensing signal, therefore enabling the CMB lensing auto-spectrum to isolate and constrain structure growth at redshifts beyond the quasar sample. We obtain a 12\% constraint on the amplitude of matter fluctuations beyond the Quaia sample, corresponding to a median redshift of $z \sim 5.1$. This measurement is consistent with predictions from primary CMB observations within 1.4$\sigma$.

This work presents one of the highest redshift measurements of structure growth, enabled by the combination of CMB lensing auto-spectrum and cross-correlation information. Future data from \textit{Gaia} DR4 are expected to improve these constraints. Also, different choices at the catalog level could be optimized for the cross-correlation measurement, such as fitting the selection functions of the Quaia samples only to the objects in the analysis footprint of the CMB lensing maps used. Upcoming galaxy survey experiments such as Euclid \citep{Euclid2025}, and the Vera C. Rubin Observatory \citep{lsstsciencecollaboration2009lsstsciencebookversion}, as well as new CMB data from the Simons Observatory \citep{thesimonsobservatorycollaboration2025simonsobservatorysciencegoals}, CMB-S4 \citep{abazajian2016cmbs4sciencebookedition}, or CMB-HD \citep{sehgal2019cmbhdultradeephighresolutionmillimeterwave}, offer promising opportunities to extend these 3$\times$2pt measurements, improving our understanding of the growth of structure and testing the $\Lambda$CDM model at high redshifts.

\section*{Acknowledgements}

We thank Mathew Madhavacheril and Neelima Sehgal for helpful discussions and comments throughout the course of this project. We also thank Roger de Belsunce for kindly making the results of their recent DESI quasar–\textit{Planck} CMB lensing cross-correlation analysis available to us.

CEV acknowledges support from “la Caixa” Foundation postgraduate fellowship (ID 100010434) with code LCF/BQ/EU22/11930099, the Wolfson College-CDT Studentship for Women in the Physical Sciences, and the Cambridge International Trust. 

BDS, CEV and GF acknowledge support from the European Research Council (ERC) under the European Union’s Horizon 2020 research and innovation programme (Grant agreement No. 851274). Computing was performed using resources provided through the STFC DiRAC Cosmos Consortium and hosted at the Cambridge Service for Data Driven Discovery (CSD3).

IAC acknowledges support from Fundaci\'on Mauricio y Carlota Botton and the Cambridge International Trust. AC acknowledges support from the STFC (grant number ST/W000977/1). JD acknowledge support from NSF grant AST-2108126, from a Royal Society Wolfson Visiting Fellowship and from the Kavli Institute for Cosmology Cambridge and the Institute of Astronomy, Cambridge. GF also acknowledges the support of the STFC Ernest Rutherford fellowship. KM acknowledges support from the National Research Foundation of South Africa. CS acknowledges support from the Agencia Nacional de Investigaci\'on y Desarrollo (ANID) through Basal project FB210003.

Support for ACT was through the U.S.~National Science Foundation through awards AST-0408698, AST-0965625, and AST-1440226 for the ACT project, as well as awards PHY-0355328, PHY-0855887 and PHY-1214379. Funding was also provided by Princeton University, the University of Pennsylvania, and a Canada Foundation for Innovation (CFI) award to UBC. The development of multichroic detectors and lenses was supported by NASA grants NNX13AE56G and NNX14AB58G. Detector research at NIST was supported by the NIST Innovations in Measurement Science program. 
ACT operated in the Parque Astron\'omico Atacama in northern Chile under the auspices of the Agencia Nacional de Investigaci\'on y Desarrollo (ANID). We thank the Republic of Chile for hosting ACT in the northern Atacama, and the local indigenous Licanantay communities whom we follow in observing and learning from the night sky.

\newpage

\bibliography{main}{}
\bibliographystyle{aasjournal}

%

\vspace{5mm}




\appendix

\section{Planck PR4 lensing $\times$ Quaia reanalysis} \label{app:Planck-reanalysis}

In this section, we outline the main differences between our analysis and previous Quaia cross-correlation studies using the \textit{Planck} PR4 CMB lensing \citep{alonso2023,Piccirilli:2024xgo}. The first modification involves minor updates to the catalogs and selection functions, already implemented and described in \citet{Piccirilli:2024xgo}. Additionally, for the covariance matrix, we employ the simulations introduced in Section~\ref{sec:simulations} and the methodology detailed in Section~\ref{sec:covariance}, in contrast to the analytical Gaussian covariance used in the previous analyses.

A key change in our baseline analysis is the treatment of the stochastic noise amplitude. Whereas the previous analyses fixed it to the value of the shot noise measured from the catalog using Equation~\eqref{eq:sn}, we adopt a Gaussian prior centered on that value with a 10\% width. This choice is motivated by the fact that variations in the stochastic noise component can significantly impact the inferred cosmological parameters, particularly $\sigma_8$, since it modulates the overall amplitude of the $C_\ell^{gg}$ signal. Given that the sample is shot-noise dominated across all scales -- with shot noise levels approximately 5--8 times larger than the signal -- our results are especially sensitive to any mis-estimation of this component, such as those arising from halo exclusion.

When performing parameter-level consistency tests, we find that our measurement is not robust to changes in scale cuts when the shot noise is allowed to vary, contrary to the findings in the previous analysis where the shot noise was held fixed. This was the main justification for pushing to more non-linear scales (up to $k_{\text{max}} = 0.22\,h\text{Mpc}^{-1}$) in the previous analysis, since negligible shifts were found in the inferred cosmological parameters between $k_{\text{max}} = 0.15\,h\text{Mpc}^{-1}$ and $k_{\text{max}} = 0.3\,h\text{Mpc}^{-1}$. However, this stability no longer holds when the shot-noise amplitude is treated as a free parameter, as we observe significant parameter shifts as the scale cuts vary. As a result, we adopt a more conservative choice of $k_{\text{max}} = 0.15\,h\text{Mpc}^{-1}$, which is also more consistent with the use of a linear galaxy bias model.

We also explore the impact of these choices on the ACT footprint using \textit{Planck} data. For Bin 2, we find a best-fit shot noise amplitude of $A_{N2} = 1.0115 \pm 0.0067$ under our fiducial $k_{\text{max}} = 0.15 \,h\text{Mpc}^{-1}$ scale cut. Fixing this amplitude instead (i.e., setting $A_{N2} = 1$) results in a roughly $0.9\sigma$ shift in $\sigma_8$, highlighting how a 1\% variation in shot noise can affect cosmological constraints. Bin 1, in contrast, appears more stable under these changes, and has an inferred amplitude of $A_{N1} = 1.001 \pm 0.014$.

\begin{figure*}
 \centering
 \includegraphics[width=\linewidth]{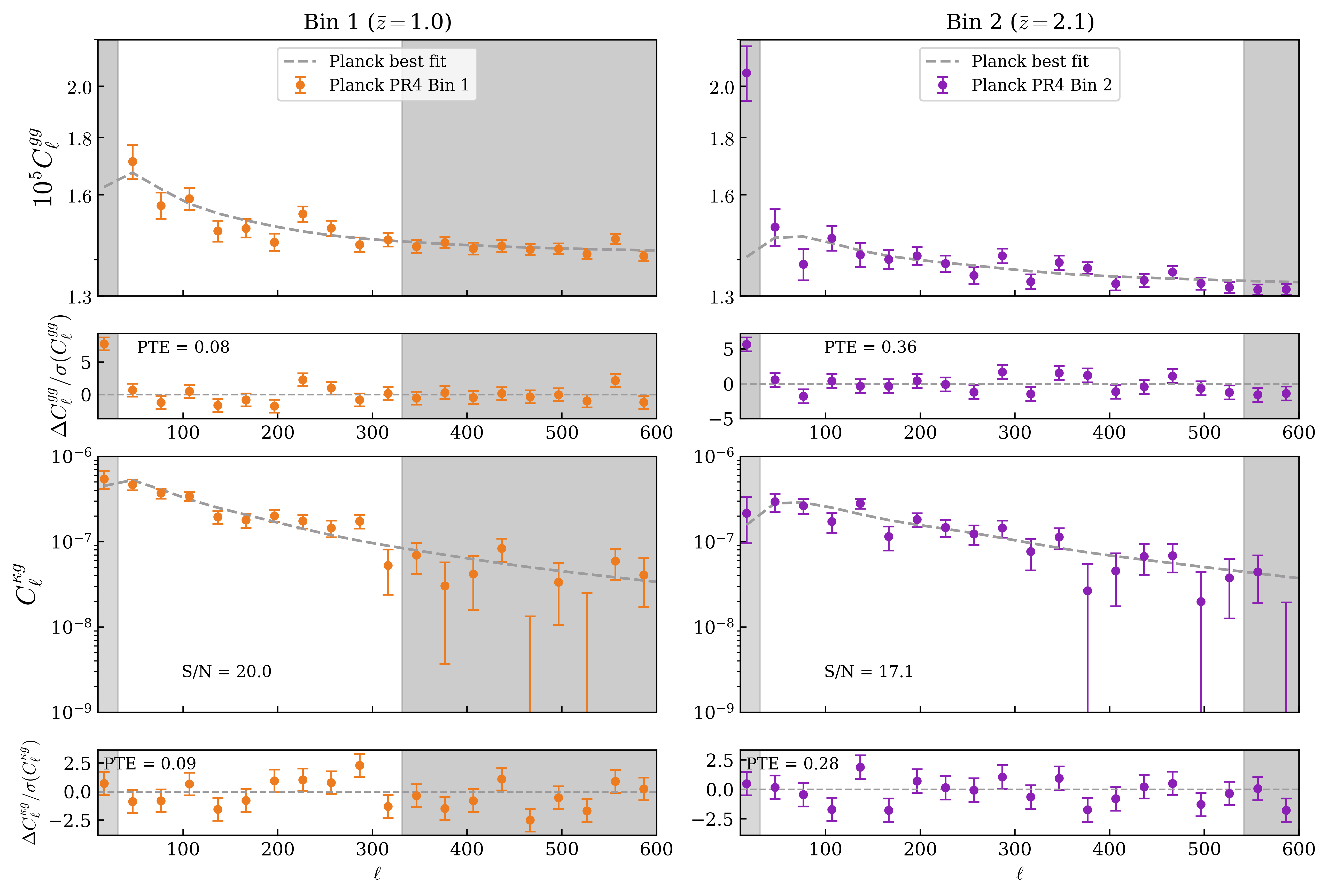}
 \caption{Measurements of $C_\ell^{gg}$ on the full footprint (top) and $C_\ell^{\kappa g}$ (bottom) using \textit{Planck} lensing for both redshift bins considered. Dashed gray lines indicate the best-fit model from the joint fit to both redshift bins. The cross-correlation with each of the redshift bins is detected with SNRs of 20.0 and 17.1 for Bin 1 and Bin 2, respectively, within the analysis range. Panels below each measurement show the model residuals.}
 \label{fig:measurement_Planck}
\end{figure*}

These two changes -- the inclusion of free shot-noise amplitudes and the more conservative scale cuts -- are the primary reasons for the degradation of our parameter constraints relative to earlier results. Nonetheless, we consider these choices more appropriate for the Quaia sample and the use of a linear bias model, and therefore apply them across both ACT and \textit{Planck} lensing cross-correlation analyses. Figure~\ref{fig:measurement_Planck} shows the updated measurements of $C_\ell^{gg}$ and $C_\ell^{\kappa g}$, along with the best-fit model jointly fit to both redshift bins. We obtain a value for our best-constrained parameter $S_8^{\times}=\sigma_8(\Omega_m/0.3)^{0.4} = 0.869^{+0.053}_{-0.075}$, and when including BAO data we directly constrain the amplitude of matter fluctuations to be $\sigma_8 = 0.797^{+0.048}_{-0.063}$.

\section{Quaia HOD}\label{app:HOD}

A halo occupation distribution (HOD) provides a framework to relate statistically the virial mass $M$ of a dark matter halo to the number of galaxies $N$ it hosts \citep{Berlind_2002}. Typically, a HOD separates contributions from central galaxies and satellite galaxies, specifying:
\begin{itemize}
    \item $N_{\text{c}}$ -- the number of central galaxies, which is either 0 or 1 depending on a minimum-mass threshold $M_{\text{min}}$.
    \item $N_{\text{s}}$ -- the number of satellite galaxies, usually assumed to be randomly distributed along a radial profile of the dark matter halo \citep[e.g.,][]{NFW_profile}.
\end{itemize}
For the purpose of populating the \texttt{WebSky} simulations as described in Section \ref{sec:tests-simbased}, we adopt a simple HOD that describes the Quaia sample. In particular, we assume that each halo above a minimum mass $M_{\text{min}}$ hosts exactly one central quasar and no satellites. This is motivated by the low number density of the sample, making the contribution from satellite quasars negligible.

Given this prescription we can relate the measured effective linear bias parameter, $b_{\text{eff},i}$, of each of the Quaia samples $i$ to the underlying halo population via
\begin{equation}
    b_{\text{eff},i} = \frac{1}{\bar{n}_g(\bar{z}_i)} \int_{M_{\text{min}}}^{M_{\text{max}}} dM \frac{dn}{dM} b(M,\bar{z}_i) \left[ N_c(M) + N_s(M) \right] \; ,
    \label{eq:befftinker}
\end{equation}
where $b(M,\bar{z}_i)$ is the halo bias as a function of the halo mass and redshift -- in our case, taken from \citet{Tinker_2010} -- and evaluated at the mean redshift of each of the samples $\bar{z}_i$, and $dn/dM$ is the halo mass function, for which we use the prescription developed in \cite{Tinker_2008}. The mean number density of galaxies is denoted $\bar{n}_g(z)$, and $N_c(M)$ and $N_s(M)$ are the average number of central and satellite galaxies per halo mass $M$, respectively.

Since our model assumes no satellite galaxies, this expression simplifies to a model in which only halos above a minimum mass $M_{\text{min}}$ host one central galaxy. We determine $M_{\text{min}}$ in each redshift bin by requiring the right-hand side of Equation~\eqref{eq:befftinker} to match the observed bias $b_{\text{eff},i}$ in each redshift bin. Since the test was performed prior to unblinding, we choose the bias values used to construct the fiducial spectra $C_{\ell\text{,fid}}^{gg}$ and $C_{\ell\text{,fid}}^{\kappa g}$. These bias values are obtained by fitting our measured spectra with a fixed cosmology (see Section \ref{sec:simulations} for further details). 

We find minimum halo masses of $M_{\text{min,}1} = 1.4\times 10^{12} \,h^{-1}M_\odot$ and $M_{\text{min,}2} = 1.3\times 10^{12}\,h^{-1}M_\odot$ for Bins 1 and 2, respectively\footnote{All halo masses in this section are defined as $M_{200\mathrm{m}}$.}. We note that the minimum mass of the \texttt{WebSky} halos in these redshifts is lower, approximately $M_{\text{min,}\texttt{WebSky}} = 9.0\times 10^{11} \, h^{-1}M_\odot$. These values are comparable to those reported in previous analyses of quasar halo masses at similar redshifts \citep{Geach_2019}, although we note that the magnitude of the two samples differ. We emphasize that this is not intended as an extensive analysis of the Quaia halo masses, but rather as a reasonable prescription for populating the \texttt{WebSky} simulations used in the extragalactic foreground contamination tests presented in Section~\ref{sec:tests-simbased}. All of these calculations were performed and are implemented in  \texttt{class\_sz}. 

\section{\texttt{WebSky} catalogs validation}\label{app:Websky}

\begin{figure*}
 \centering
 \includegraphics[width=0.7\linewidth]{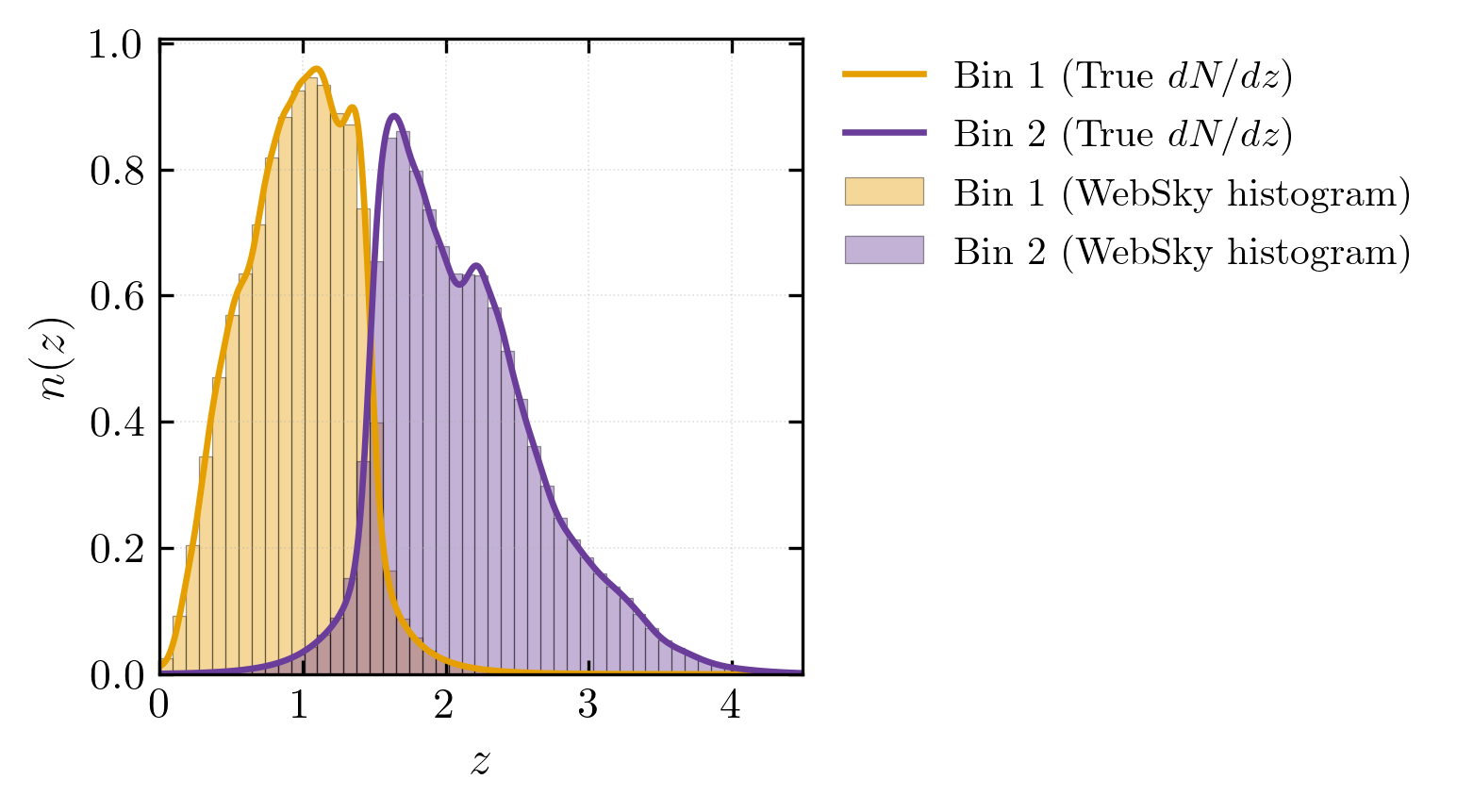}
 \caption{Comparison between the redshift distributions of the \texttt{WebSky} mock catalogs (histograms) and the true $dN/dz$ of the Quaia redshift bins (solid lines). These mocks are constructed to match both the redshift distribution and number density of the Quaia samples and are used in the extragalactic foreground tests described in Section~\ref{sec:tests-simbased}.}
 \label{fig:websky_dndz}
\end{figure*}

After determining $M_{\text{min}}$ for each redshift bin, we populate the \texttt{WebSky} halo catalog according to the HOD prescription described in Appendix \ref{app:HOD}. We then sample the resulting catalogs to match both the redshift distribution and angular number density of the Quaia samples. Figure~\ref{fig:websky_dndz} shows the redshift histograms of the final \texttt{WebSky} mock catalogs compared to the $dN/dz$ distributions of the Quaia bins.

\begin{figure*}
 \centering
 \includegraphics[width=\linewidth]{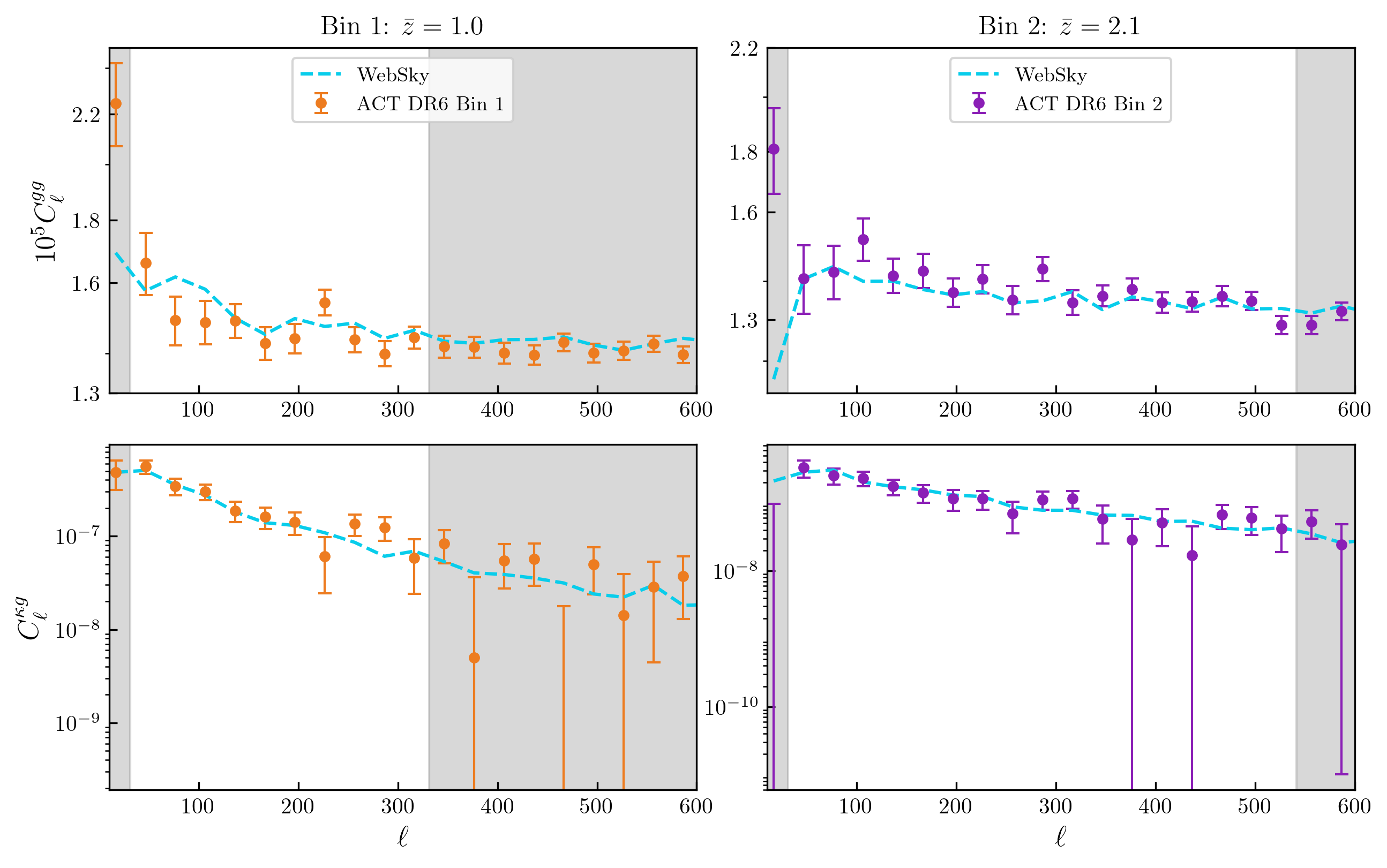}
 \caption{Validation of the final \texttt{WebSky} catalogs by comparing their auto-correlation ($C_\ell^{gg}$; top) and cross-correlation with CMB lensing ($C_\ell^{\kappa g}$; bottom) to the measured spectra in both redshift bins. This agreement supports the use of these mocks in the simulation-based foreground-contamination tests described in Section~\ref{sec:tests-simbased}.}
 \label{fig:websky_spectra}
\end{figure*}

We also validate the resulting catalogs by comparing their angular power spectra to our measurements. Specifically, we compute both the auto-correlation, $C_\ell^{gg}$, and the cross-correlation with CMB lensing, $C_\ell^{\kappa g}$, for each redshift bin. As shown in Figure~\ref{fig:websky_spectra}, the simulated spectra reproduce the overall shape and amplitude of the observed spectra, demonstrating that our simplified HOD model provides a reasonable match to the clustering properties of the Quaia sample on the scales used in our analysis.

\end{document}